\documentclass[reprint,superscriptaddress,amsmath,amssymb,aps,prl,longbibliography]{revtex4-2}
\usepackage{graphicx}
\usepackage{dcolumn}
\usepackage{bm,bbm}
\usepackage{xcolor}
\usepackage{tcolorbox}
\usepackage{algorithm}
\usepackage{algpseudocode}
\usepackage{amsmath}
\usepackage{amsmath}
\usepackage{amssymb}
\usepackage{amsfonts}
\usepackage{graphicx}
\usepackage{float}
\usepackage{subfigure}
\usepackage{color}
\usepackage{float}
\usepackage{mathrsfs}
\usepackage{amsmath}
\usepackage{bbm}
\usepackage{booktabs} 
\usepackage[normalem]{ulem}

\usepackage[citecolor=blue,colorlinks,breaklinks]{hyperref}
\makeatletter

\usepackage{xcolor}
\usepackage{lipsum}

\newcommand{\M}{M}
\definecolor{myred}{rgb}{0.53, 0, 0.08}

\newcommand{\target}{\Lambda}

\begin{document}

\preprint{APS/123-QED}

\title{Experimental High-Dimensional Quantum Overlapping Tomography}%

\author{Haifei Wang}
\affiliation{%
Centre for Quantum Technologies, Singapore 117543, Singapore}
\affiliation{%
Quantum Science and Engineering Centre (QSec), Nanyang Technological University, Singapore 639798, Singapore
}%

\author{Rui Qu}
\email{rui.qu@ntu.edu.sg}
\affiliation{%
Centre for Quantum Technologies, Singapore 117543, Singapore}
\affiliation{%
School of Electrical and Electronic Engineering, Nanyang Technological University, Singapore 639798, Singapore
}%

\author{Zhengning Yang}
\affiliation{%
Division of Physics and Applied Physics, School of Physical and Mathematical Sciences, Nanyang Technological University, Singapore, 637371, Singapore}

\author{Xiaodan Lyu}
\affiliation{%
School of Electrical and Electronic Engineering, Nanyang Technological University, Singapore 639798, Singapore
}%
\author{Haotao Zhu}
\affiliation{%
School of Electrical and Electronic Engineering, Nanyang Technological University, Singapore 639798, Singapore
}%

\author{Zitong Xu}
\affiliation{%
School of Electrical and Electronic Engineering, Nanyang Technological University, Singapore 639798, Singapore
}%

\author{Lianzhen Cao}
\affiliation{School of Physics and Electronic Information, Weifang University, Weifang 261061, China}

\author{Joel Yang}
\affiliation{Engineering Product Development, Singapore University of Technology and Design, Singapore 487372, Singapore}

\author{Otfried G\"uhne}
\email{otfried.guehne@uni-siegen.de}
\affiliation{Naturwissenschaftlich-Technische Fakult\"at, Universit\"at Siegen, Walter-Flex-Stra{\ss}e 3, 57068 Siegen, Germany}

\author{Weibo Gao}
\email{wbgao@ntu.edu.sg}
\affiliation{%
Centre for Quantum Technologies, Singapore 117543, Singapore}
\affiliation{%
Quantum Science and Engineering Centre (QSec), Nanyang Technological University, Singapore 639798, Singapore
}%
\affiliation{%
School of Electrical and Electronic Engineering, Nanyang Technological University, Singapore 639798, Singapore
}%
\affiliation{%
Division of Physics and Applied Physics, School of Physical and Mathematical Sciences, Nanyang Technological University, Singapore, 637371, Singapore}

\begin{abstract}
Large-scale quantum systems have advanced rapidly via the exploration of more particles and higher dimensions, offering great potential for developing quantum technologies.
However, their characterization becomes prohibitive with increasing local dimensionality and particle number. %
Here we propose high-dimensional quantum overlapping tomography based on a graph-theoretic formulation, which allows one to efficiently reconstruct few-body marginals of multipartite high-dimensional quantum systems. 
We experimentally realize it on a photonic four-party entangled state 
in a $4 \times 4 \times 2 \times 2$ system.
Using measurements in mutually unbiased bases,
we reconstruct all six two-body marginals with only 
25 projective measurement settings, compared with 94 and 225 settings for independent tomography of all two-body reduced states and full state tomography, respectively. 
The reconstructed marginals reveal a layered entanglement structure vital for high-dimensional quantum networks.
We further show that these marginals enable more noise-resilient certification of multipartite high-dimensional entanglement than the fidelity-based criterion. 
Our work thus offers a scalable route for learning multidimensional quantum 
systems.

\end{abstract}

\date{\today}

\maketitle

\textit{Introduction.—}Complex quantum systems involving multilevel quantum particles provide a powerful route to enlarging accessible Hilbert spaces without increasing the number of physical carriers, revealing new fundamental physical perspectives and applications in quantum information processing \cite{PhysRevLett.85.3313,PhysRevLett.88.127902,erhard2020advances, chi2022programmable,ringbauer2022universal,brock2025quantum,meth2025simulating, yu2025quantum,54mc-2yl3}. Recent advances have explored quantum entangled systems featuring both higher dimensions and multiple particles in bulk \cite{malik2016multi, erhard2018experimental, PhysRevLett.120.260502, hu2020experimental, hu2025observation} and on-chip optical platforms \cite{zheng2023multi, bao2023very}. Their enlarged local dimension enables richer entanglement structures \cite{cobucci2024detecting}, higher information capacity \cite{groblacher2006experimental,hu2018beating} and improved noise resilience for quantum information tasks \cite{PhysRevX.9.041042, PhysRevApplied.15.034003,Qu-PhysRevLett.128.240402}. However, it also makes the complete characterization of the global system substantially more demanding with increasing dimensionality and particle number \cite{cramer2010efficient, toninelli2019concepts}. 

Many quantum applications, however, require only partial characterization, for example through entanglement witnesses \cite{RevModPhys.81.865, PhysRevLett.111.110503} or local Hamiltonian terms \cite{PhysRevLett.122.020504, clinton2021hamiltonian, clinton2024towards}. 
This observation motivates efficient prediction protocols such as classical shadows \cite{huang2020predicting, PhysRevLett.127.200501, huang2022learning}. 
An important feature of most application-relevant partial information is locality, as physical Hamiltonians and noise processes are often dominated by few-particle interactions, and the relevant diagnostics are encoded in few-body correlations. 
In particular, reduced density matrices determine all few-body correlations, enter many entanglement-certification criteria \cite{friis2019entanglement} and even uniquely determine the underlying state in several generic settings \cite{PhysRevLett.89.207901, PhysRevA.70.010302, PhysRevA.96.010102, yu2023learning, zhang2024almost}. In multipartite high-dimensional systems, a central property is the entanglement dimensionality, quantified by the Schmidt number across each bipartition \cite{PhysRevLett.110.030501}. It is commonly certified through the fidelity with a target state \cite{malik2016multi, erhard2018experimental, hu2020experimental, hu2025observation, bao2023very}, but criteria that combine the fidelity with reduced density matrices detect it in more cases \cite{PRXQuantum.2.040357, li2026simultaneous}. Such criteria require many few-body marginals. 	

Quantum overlapping tomography (QOT) is an efficient method to extract such few-body information.
Rather than measuring each $k$-partite subsystem independently, QOT enables the reconstruction of all $k$-partite marginals with a substantially smaller number of settings by avoiding information overlap among the set of global measurements \cite{cotler2020quantum, PhysRevResearch.2.023393, PhysRevX.10.031064}. 
For reconstructing all $k$-body marginals in $n$-qubit systems, the original protocol \cite{cotler2020quantum} reduces Pauli settings from the scaling of $e^{\mathcal{O}(k)} n^k$ to $e^{\mathcal{O}(k)}\log n$ with perfect hash functions. It was then experimentally demonstrated in photonic \cite{yang2023experimental} and superconducting systems \cite{PhysRevLett.133.160801}, and used in nuclear-spin systems \cite{qin2024experimental, wei2024optimal, xu2026multiparticle}. 
In particular, recent work demonstrated that the optimal scheme of QOT can be formulated using graph-theoretic constructions \cite{wei2024optimal} and covering arrays \cite{colbourn2004combinatorial, mixedca, sarkar2017upper}, yielding a further reduction in the number of measurements \cite{hansenne2024optimal},
and allowing for a generalization to fermionic systems \cite{ghoshal2025qubits} and to qudits measured with generalized Gell-Mann matrices, which require $d^2-1$ local settings for a qudit of dimension $d$ \cite{ma2026optimal}.
However, existing qudit formulations of QOT \cite{hansenne2024optimal, ma2026optimal} use more than the minimal $d+1$ local projective settings per qudit for $d>2$ and assume a common local dimension. The experimental realization of high-dimensional QOT also remains unexplored.

In this work, we propose QOT for general high-dimensional systems based on a graph-theoretic formulation with $d+1$ local settings per qudit and arbitrary local dimensions. We then experimentally demonstrate it on a photonic four-partite hybrid qudit--qubit entangled state with local dimensions $(4,4,2,2)$. 
With only 25 mutually unbiased basis (MUB) settings, we reconstruct all two-body marginals with an average fidelity of $0.876 \pm 0.014$ with respect to the target marginals. The reconstruction reveals 
the intended asymmetric pairwise entanglement structure of the prepared state, useful for layered quantum communication in quantum networks.
Moreover, we show that these marginals provide a noise-resilient route to certifying the entanglement dimensionality vector \cite{PhysRevLett.110.030501} of multipartite quantum systems, as they supply the local overlaps that the inner-product criterion combines with the global fidelity.
Our work thus provides an efficient and scalable characterization framework for large-scale high-dimensional quantum systems, in the sense that, for fixed local dimension and marginal size, the number of measurement settings and the sample complexity grow only logarithmically with the number of parties.

\textit{Protocol.—}We first exploit high-dimensional QOT based on graph theory for reconstructing all two-body marginals of the global quantum system.
For a prime-power dimension $d$, a complete set of MUBs contains $(d+1)$ orthonormal bases, 
with the property $|\langle \psi | \phi \rangle|^2=1/d$ for any vector $|\psi\rangle$ and $|\phi\rangle$ from two different bases \cite{wootters1989optimal}. Quantum measurements based on MUBs form an informationally complete set for single-qudit tomography \cite{lima2011experimental, PhysRevLett.110.143601}. 
For non-prime-power dimensions, informationally complete sets of $(d+1)$ projective measurement settings can also be constructed, though with less symmetry than the MUBs \cite{xiao2025d}. 
For simplicity, we restrict the following discussion to prime-power dimensions and MUB measurement settings.

\begin{figure}[t!]
    \centering
    \includegraphics[width=.9\linewidth]{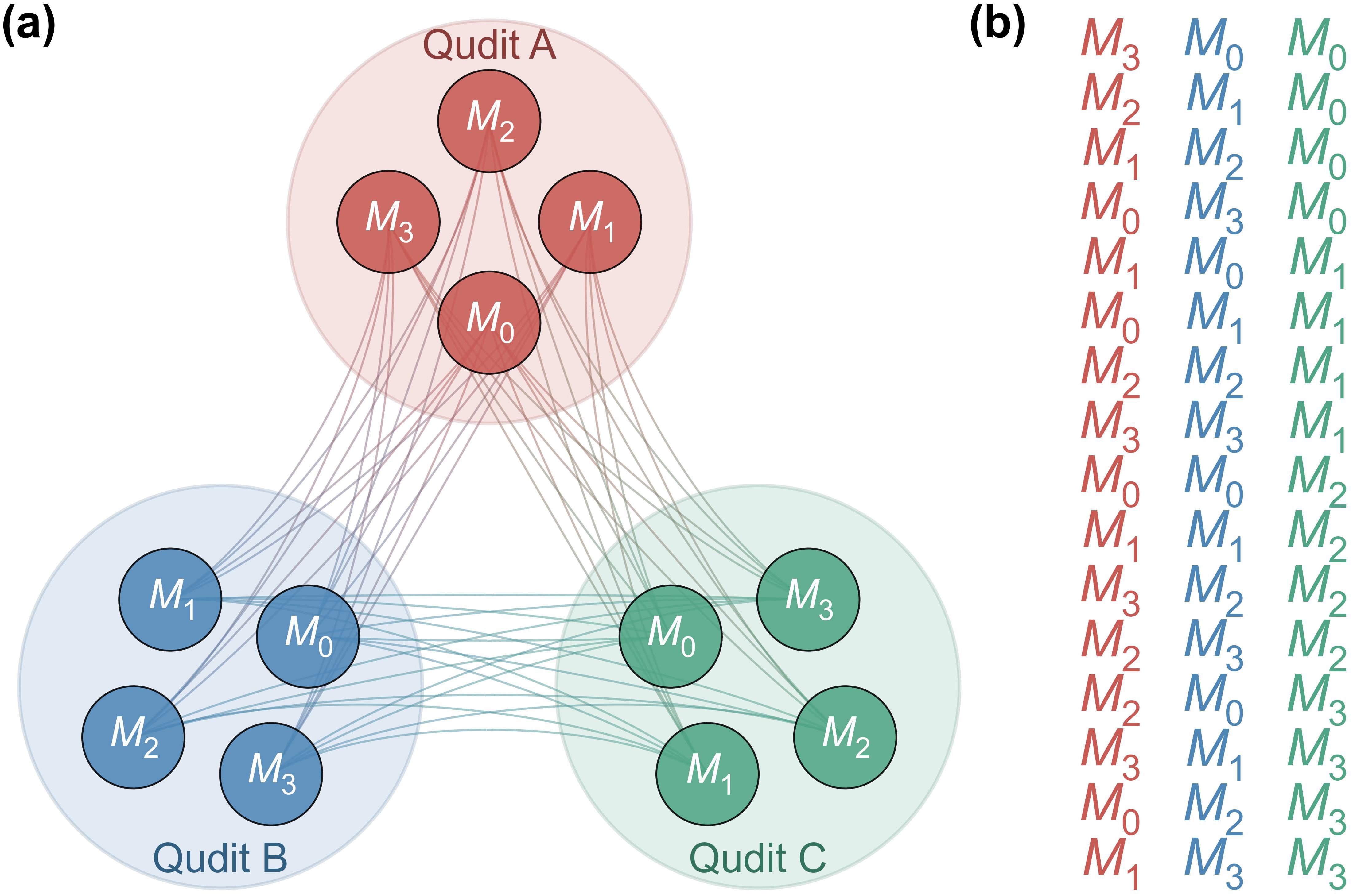}
    \caption{\textbf{Schematic for high-dimensional overlapping tomography.} (a) Complete multipartite graph $K_{4,4,4}$ for three qutrits, where each qutrit is associated with four vertices representing its MUB settings $\M_{0},\M_{1},\M_{2},\M_{3}$. Edges represent tensor products of two MUB projectors. Cliques (triangles) correspond to three-body measurement settings (tensor products of three MUB projectors). (b) The 16 cliques that cover all 48 edges of the graph. The corresponding MUB settings are thus informationally complete for extracting all two-body marginals.}
    \label{fig:figure_1}
\end{figure}

We start the procedure by constructing a ``complete multipartite graph'' \cite{chartrand2019chromatic}.
For each particle with local dimension $d_i$, we let $(d_i+1)$ vertices represent its complete MUB settings $\M_0,\M_1,\ldots,\M_{d_i}$. 
We then add edges between every pair of vertices belonging to different particles, forming the complete multipartite graph
$K_{\vec{D}}:=K_{d_1+1,d_2+1,\ldots,d_n+1}$. A global measurement setting is represented by choosing one vertex from each particle and connecting all of them, forming a complete subgraph, or a \textit{clique}. %
An example of $K_{4,4,4}$ is shown in Fig.~\ref{fig:figure_1}, corresponding to a system consisting of three qutrits with $d_i=3$. 
In this graph, each clique contains three edges, which means that a single global measurement setting simultaneously covers three two-body measurement settings. 
Consequently, the optimal QOT protocol for reconstructing all two-body marginals is equivalent to covering all edges of $K_{\vec{D}}$ with the smallest possible set of cliques \cite{hansenne2024optimal, wei2024optimal}. 
For $K_{d+1,d+1,d+1}$, only $(d+1)^2$ measurement settings are actually needed to reconstruct all two-body marginals.
To give concrete numbers, for $K_{4,4,4}$ 
only 16 measurement settings are needed while 
independent full state tomography on each of the two-body marginals would require $3\times 16 = 48$ settings.

Let us illustrate the scalability of high-dimensional QOT.
Considering an $n$-partite system with the same local dimension $d$ in each party, the relevant graph is the complete multipartite graph $K_{\vec{D}}$, where $\vec{D}=(d+1,...,d+1)$ with $(d+1)$ appearing $n$ times.
The corresponding minimum number of cliques for $k$-partite marginals is denoted by $\phi_k(n,d)$. 
For fixed $d$, $\phi_k(n,d)$ grows only logarithmically with $n$, scaling as $(d+1)^k\mathcal{O}(\log n)$, as demonstrated in Sec.~S1\,B of the Supplemental Material. Thus, the number of settings remains far less than $(d+1)^k\binom{n}{k}$ settings required to measure every $k$-body marginal independently. For $k>2$, the graph formulation generalizes naturally to covering arrays; details of the general construction are given in Sec.~S1\,A of the Supplemental Material.

\begin{figure}[th!]
    \centering
    \includegraphics[width=\linewidth]{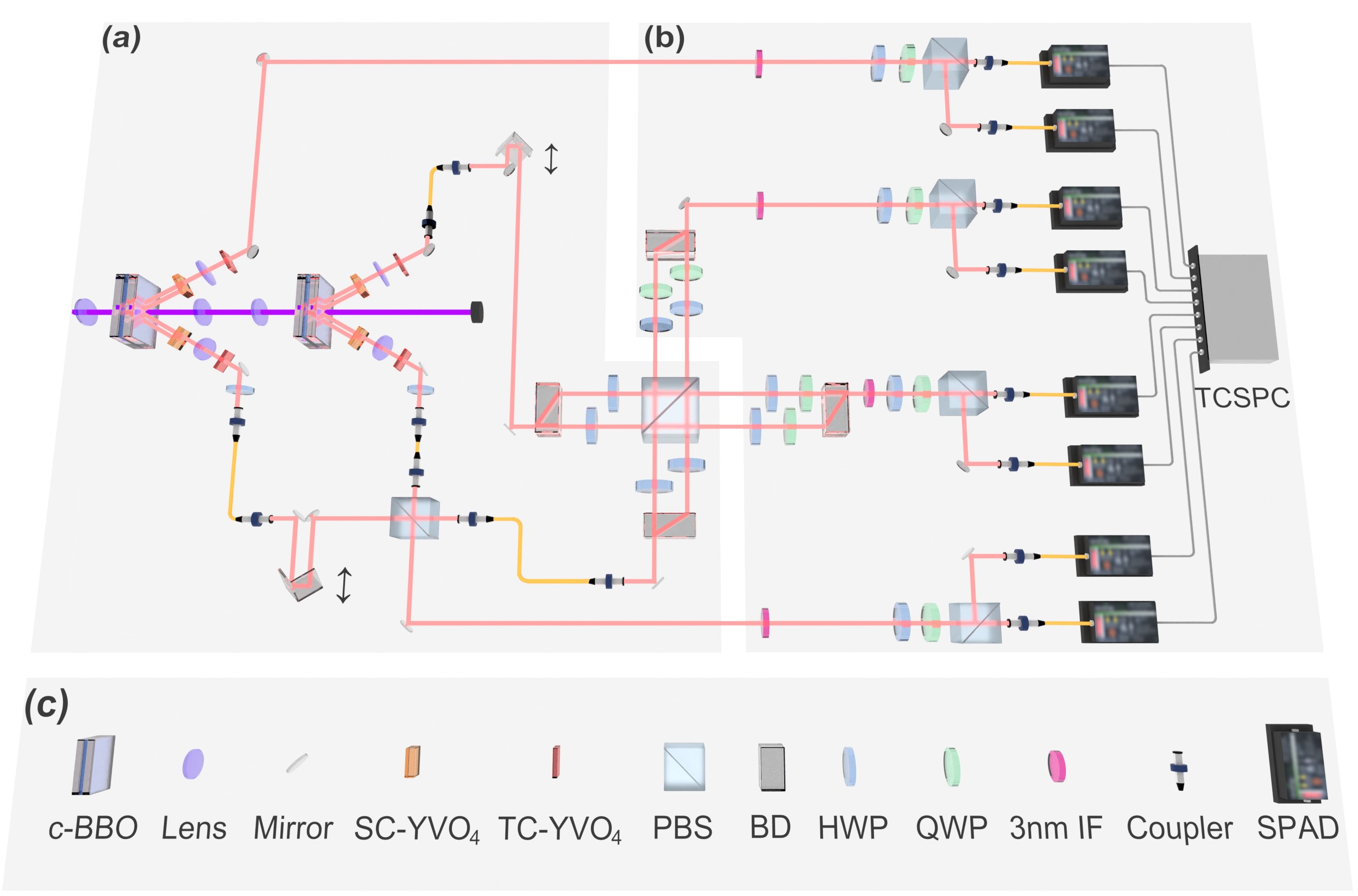}
    \caption{
    Experimental setup.
    (a) Preparation of the target state $|\target\rangle$. 
    Ultraviolet femtosecond pulses (390~nm, 80~MHz repetition rate, 230~mW average power) are focused onto two sandwich-like BBO--HWP--BBO combinations (c-BBO) to generate two polarization-entangled photon pairs at 780~nm via spontaneous parametric down-conversion. 
    Spatial and temporal walk-off in the BBO crystals is compensated by yttrium orthovanadate (YVO$_4$) crystals. The BDs displace vertically polarized photons into the upper path while leaving horizontally polarized photons in the lower path. 
    Narrow-band interference filters (IF) with a 3~nm passband are used to suppress unwanted frequency correlations between the signal and idler photons. 
    By postselecting fourfold coincidence events, 
    the target state is generated after the second PBS.
    (b) Measurement of the target state. 
    The two ququarts are analysed by measurement modules composed of BDs, wave plates, PBSs, and single-photon avalanche diodes (SPADs), allowing projections onto MUB settings. 
    Fourfold coincidence events are recorded by a time-correlated single-photon counting (TCSPC) module.
    (c) Optical elements used in the setup. BBO: $\beta$-barium borate; SC-YVO$_4$: spatial-compensation YVO$_4$; TC-YVO$_4$: temporal-compensation YVO$_4$; BD: beam displacer; HWP: half-wave plate; QWP: quarter-wave plate.}
    \label{fig:figure_2}
\end{figure}

\textit{Experimental demonstration.—}Experimentally, we prepare a photonic four-partite entangled state
\begin{equation}
    |\target\rangle = \frac{1}{2}\left(
    |0011\rangle + |1100\rangle + |2211\rangle + |3300\rangle
    \right),
    \label{eq:target_state}
\end{equation}
where two parties are ququarts and the other two parties are qubits.
In particular, this state has nonuniform entanglement dimensionality across different bipartitions. 
For the bipartitions of one vs.\ three parties 
($1|3$ cuts), the elements of the Schmidt number vector \cite{PhysRevLett.110.030501} are
\begin{equation}
    [r_{A|BCD}\ r_{B|ACD}\ r_{C|ABD}\ r_{D|ABC}]
    =
    [4\ 4\ 2\ 2],
\end{equation}
while for the $2|2$ bipartitions, they are
\begin{equation}
    [r_{AC|BD}\ r_{AD|BC}\ r_{AB|CD}]
    =
    [4\ 4\ 2].
\end{equation}
In particular, one can remotely distribute the Bell 
state $(|11\rangle+|33\rangle)_{AB}$ between the 
parties $A$ and $B$ by projecting $C$ and $D$ onto 
$(|00\rangle)_{CD}$, while projecting onto 
$(|11\rangle)_{CD}$ prepares a Bell state
$(|00\rangle+|22\rangle)_{AB}$ on $A$ and $B$, 
acting on different levels of the high-dimensional
quantum system. 
In this sense, the high-dimensional 
entanglement of the state $|\target\rangle$ leads to 
a flexibility that can exhibit advantages in constructing 
complex quantum networks \cite{hu2020experimental}. This 
also makes it an interesting test-bed for demonstrating high-dimensional QOT.

Our experimental setup for demonstrating high-dimensional QOT is depicted in Fig.~\ref{fig:figure_2}. 
We first generate two polarization Bell pairs $|\Psi^+\rangle = (|HH\rangle + |VV\rangle)/\sqrt{2}$ via spontaneous parametric down-conversion in $\beta$-barium borate crystals, where $H$ and $V$ denote the horizontal and vertical polarization states. 
We then interfere them with a polarizing beam splitter (PBS) and a high-dimensional interferometric device consisting of a PBS and two beam displacers (BDs). 
The BDs displace vertically polarized photons into the upper path by 1.3~mm while leaving horizontally polarized photons in the lower path, that is, $|H\rangle \rightarrow |H_l\rangle$ and $|V\rangle \rightarrow |V_u\rangle$. 
We encode the hybrid path--polarization states as $|H_u\rangle \rightarrow |0\rangle$, $|H_l\rangle \rightarrow |1\rangle$, $|V_u\rangle \rightarrow |2\rangle$, and $|V_l\rangle \rightarrow |3\rangle$. 
The target state $|\target\rangle$ is generated with half-wave plates placed at $22.5^\circ$ after the BDs. 
With temporal delay control using retroreflectors and 3~nm narrow-band filtering to separate unwanted frequency correlations, we obtain a four-photon coincidence rate of 0.8~Hz with an interference visibility of $0.893\pm0.005$. 

To characterize the experimentally generated state, we first 
determine its fidelity $F=\langle \target | \rho |\target\rangle$ 
with the target state by evaluating 35 measurement settings; 
the settings are detailed in Sec.~S2\,B of the Supplemental Material. 
From these data, we determine the experimental fidelity to be 
$0.809 \pm 0.009$. The best achievable overlap between any 
state which has smaller Schmidt number (for at least one 
of the bipartitions) and the ideal state 
$|\target\rangle\langle\target|$ is $F_{\max} = 0.75$ (see Supplemental Material, Sec.~S2\,E). Thus, our state is proved to be entangled with a 
Schmidt number vector of
\begin{align}   
    d_{\rho}=&[r_{A|BCD}\ r_{B|ACD}\ r_{C|ABD}\ r_{D|ABC}\ r_{AC|BD}\ \nonumber\\&r_{AD|BC}\ r_{AB|CD}]
    =[4\ 4\ 2\ 2\ 4\ 4\ 2].
\end{align}
For two-body high-dimensional QOT, we collect coincidence counts 
for 600 seconds under each setting. We repeat this process for 
the 25 measurement settings required to reconstruct all six 
two-body marginals of the global quantum system. These measurement settings are obtained by solving the clique cover problem for the complete multipartite graph $K_{5,5,3,3}$. The measurement settings (Table~S1), waveplate configurations (Table~S3), relevant experimental data (Sec.~S2\,D), and confidence region analysis \cite{PhysRevA.109.062417} (Sec.~S1\,C) are detailed in the Supplemental Material.

\begin{figure}[t!]
    \centering
    \includegraphics[width=1\linewidth]{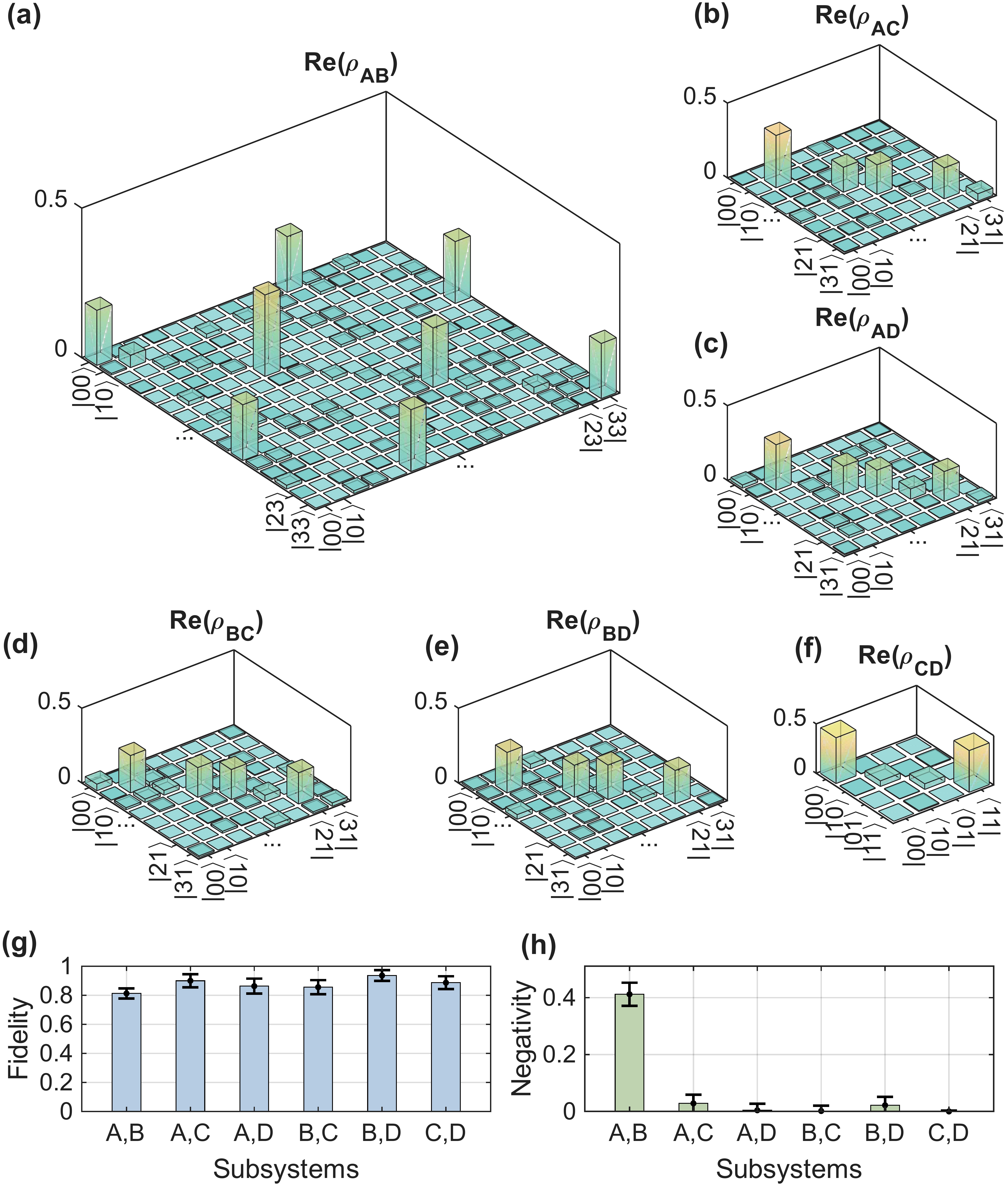}
    \caption{Experimental results of high-dimensional overlapping tomography. (a-f) Real parts of the six two-body marginals reconstructed by high-dimensional overlapping tomography. (g) The fidelities of the six two-body marginals. Error bars are obtained by Monte Carlo sampling with 1000 samples following Poisson distribution and represent 99\% confidence intervals. (h) Negativities of the two-body marginals. Only $\rho_{AB}$ has a non-zero negativity with statistical significance, indicating its entanglement, whereas the other two-body marginals do not.
    }
    \label{fig:figure_3}
\end{figure}

The real parts of the six two-body marginals reconstructed from QOT are shown in Fig.~\ref{fig:figure_3}a--f, while the imaginary parts are given in the Supplemental Material (Fig.~S5). Maximum likelihood estimation is used to ensure positivity of the reconstructed states \cite{PhysRevA.64.052312}. 
For these marginals, we obtain an average fidelity of $0.876 \pm 0.014$, with $F_{AB} = 0.812 \pm 0.012$, $F_{AC} = 0.900 \pm 0.015$, $F_{AD} = 0.863 \pm 0.017$, $F_{BC}=0.856\pm0.015$, $F_{BD}=0.936\pm0.012$, $F_{CD}=0.887\pm0.014$, respectively, as shown in Fig.~\ref{fig:figure_3}g. 
Here $F_{X}:=F(\rho_{X},\sigma_{X}) = \operatorname{tr}\left(\sqrt{\sqrt{\sigma_X}\rho_X\sqrt{\sigma_X}}\right)^2$ denotes the fidelity between the reconstructed marginal $\rho_{X}$ and the corresponding ideal marginal $\sigma_{X}$. 
Hence, high-dimensional QOT allows us to reconstruct all two-body marginals of this global state with only $25$ MUB measurement settings. For comparison, independent full state tomography of all the marginals 
would require {$5^2+4\times(5\times 3)+3^2=94$} MUB settings and full state tomography of the global state would require {$5\times 5\times 3\times 3=225$} MUB settings. For each marginal, the 25 settings contain all local basis combinations required for its independent tomography, for example all $5\times5$ combinations for $\rho_{AB}$ (Table~S1). The reduction from 94 to 25 settings therefore comes from using each global setting for all six marginals at once.

Note that the deviations from unit fidelity are mainly due to imperfections in state preparation. Among the six marginals, $\rho_{AB}$ has the lowest fidelity. Its ideal state has coherences in the $\{|00\rangle,|22\rangle\}$ and $\{|11\rangle,|33\rangle\}$ subspaces, while the other five ideal marginals are diagonal. Population outside the supports of the target marginals affects all six marginals, whereas imperfect interference further lowers the $AB$ fidelity by suppressing these coherences (Sec.~S2\,G of the Supplemental Material).

Among all reconstructed marginals, only $\rho_{AB}$ shows significant non-zero off-diagonal terms, while the other marginals have only diagonal terms. 
The marginal $\rho_{AB}$ exhibits a special correlation structure, which is a classical mixture of two Bell pairs embedded in two different orthogonal subspaces, $(|00\rangle + |22\rangle)$ and $(|11\rangle+|33\rangle)$. This implies that while it is possible to share secret keys across all four parties, the two parties $A$ and $B$ can share additional secret keys regardless of measurements by $C$ and $D$.
This structure may be useful for a layered quantum communication network, where parties with higher security levels (e.g., banks) can share extra secret keys inaccessible to parties with lower security levels (e.g., ATMs) \cite{hu2020experimental}. 
In such applications, high-dimensional pairwise QOT is particularly well suited as it can certify the security levels across every pair of parties simultaneously with efficient measurement overhead.
We can further analyze this quantitatively by calculating the entanglement monotones of the reconstructed marginals $\rho_{X}$, such as the negativity $\mathcal{N}:=\sum_{\lambda_k<0}|\lambda_k|$ where $\lambda_k$ are the eigenvalues of the partial transpose of $\rho_{X}$.
Based on the experimentally reconstructed marginals, we can obtain the negativities of each two-body subsystem, as given in Fig.~\ref{fig:figure_3}h. 
The results show that $\rho_{AB}$ is entangled with statistical significance, whereas the other two-body marginals have only classical correlations.

\begin{figure}[t!]
    \centering
    \includegraphics[width=.95\linewidth]{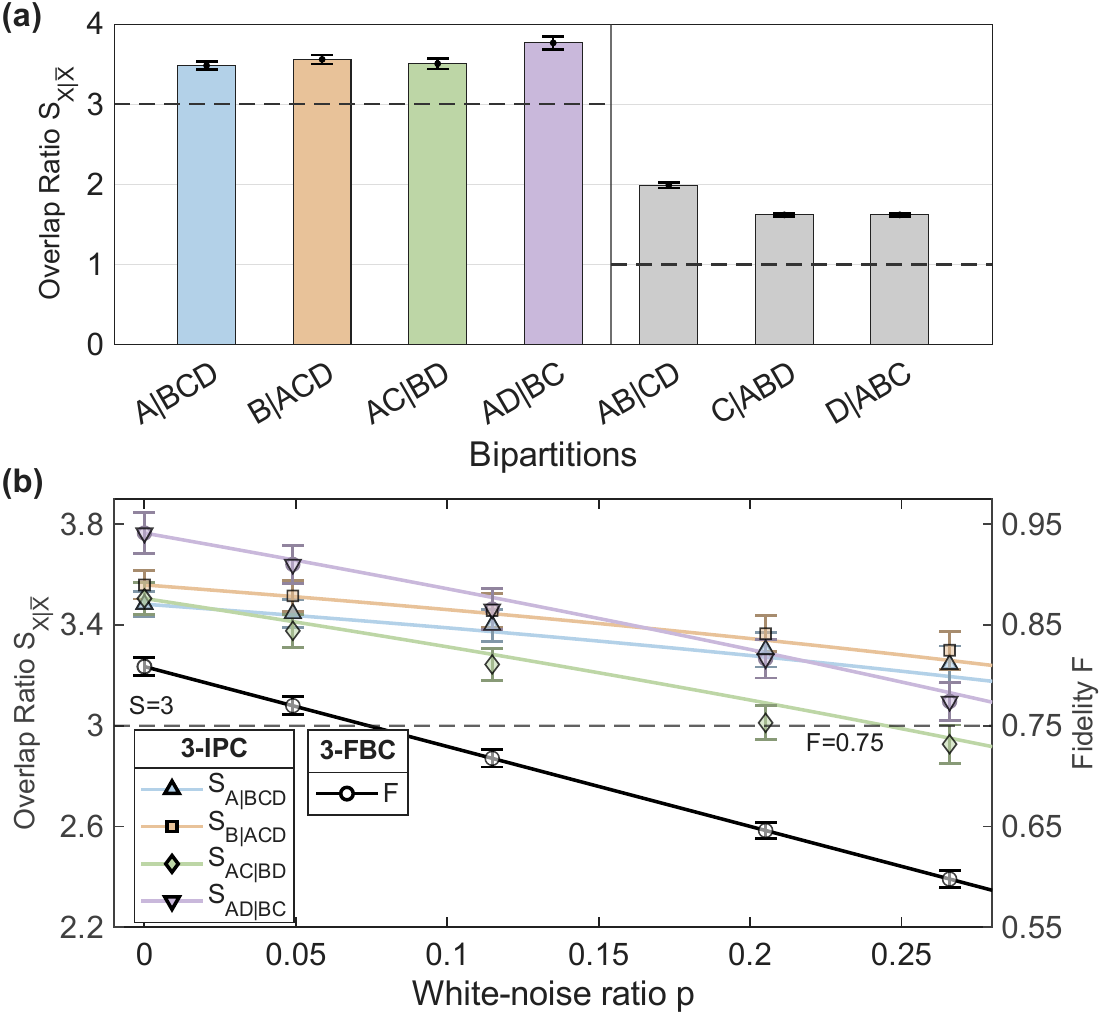}
    \caption{Certification of the entanglement dimensionality vector. (a) The global-to-local overlap ratios $S$ for all seven bipartitions. Error bars represent one standard deviation. 
    These values lower-bound the Schmidt number vector of our experimental state as $r_{A|BCD}, r_{B|ACD}, r_{AC|BD}, r_{AD|BC} > 3$ and $ r_{AB|CD}, r_{C|ABD}, r_{D|ABC} >1$.
    (b) Comparison between the $r$-inner-product criterion ($r$-IPC) and the $r$-fidelity-based criterion ($r$-FBC) for detecting the Schmidt number greater than $r$, under different levels of white noise. 
    Here $p$ is the ratio of the additional white noise sample counts to the total sample counts, obtained from postprocessing. The dashed line marks both thresholds, $S=3$ on the left axis and $F=0.75$ on the right axis. 
    }
    \label{fig:figure_4}
\end{figure}

\textit{Noise-resilient certification of entanglement dimensionality.—}The marginals obtained from QOT enable a more noise-resilient certification of the entanglement dimensionality vector. The inner-product criterion (IPC) combines these marginals with the global fidelity to lower-bound the Schmidt numbers, whereas the fidelity-based criterion (FBC) applied above relies on the fidelity alone. The IPC in theory has stronger detection power than the FBC in most cases \cite{PRXQuantum.2.040357, li2026simultaneous}.
For a subsystem $X$ of a bipartition, the global-to-local overlap ratio between two states is defined as: 
$S_X(\rho, \sigma) = \langle \rho, \sigma\rangle/\langle \rho_X, \sigma_X\rangle$ if $\langle \rho_X, \sigma_X\rangle\ne0$ and $S_X(\rho, \sigma) = 0$ otherwise. Here $\langle \rho, \sigma\rangle=\operatorname{tr}(\rho \sigma)$ is the Hilbert-Schmidt inner product. 
The Schmidt number $r_{X|\bar{X}}$ for the bipartition $X|\bar{X}$ can be lower-bounded by $r_{X|\bar{X}} \ge \lceil S_{X|\bar{X}}(\rho,\sigma) \rceil$ where $S_{X|\bar{X}}(\rho,\sigma):= \max \{S_{X}(\rho,\sigma), S_{\bar{X}}(\rho,\sigma)\}$. 
As shown in Fig.~\ref{fig:figure_4}(a), by setting $\sigma=|\target\rangle\langle \target|$ as our target state, $S_{X|\bar{X}}(\rho,\sigma)$ can be calculated from the fidelity $\langle \target|\rho|\target\rangle=\operatorname{tr}(\rho\sigma)$ and the marginals obtained by high-dimensional QOT.
For the $2|2$ bipartitions, the two-body local overlaps $\langle \rho_X, \sigma_X\rangle$ are accessible directly from the reconstructed two-body marginals $\rho_X$. For the $1|3$ bipartitions, the one-body local overlaps are available from reduced one-body marginals, 
and the three-body local overlaps for subsystems $BCD$ and $ACD$ are accessible from the computational basis measurement because the corresponding target marginals are diagonal in the computational basis; the three-body overlaps for subsystems $ABD$ and $ABC$ are not accessible and we thus only take the corresponding one-body overlap ratios $S_{C}$ and $S_{D}$. For the target state mixed with white noise, the IPC certifies Schmidt numbers of at least four up to noise ratios of 0.410 and 0.593 for the $2|2$ and $1|3$ bipartitions, compared with 0.254 for the FBC (Sec.~S2\,E and Fig.~S6 of the Supplemental Material).

To experimentally verify the noise robustness, we evaluate the two criteria for different levels of environmental noise. 
This is achieved by putting independent noise sources in front of each optical coupler to introduce additional white noise; 
see Sec.~S2\,F of the Supplemental Material for details.
Thus, the noise model can be expressed as $\rho(p) = (1-p)\rho + p\mathbb{I}/64$, where $\mathbb{I}$ is the identity matrix in $\mathbb{C}^{64}$.
As shown in Fig.~\ref{fig:figure_4}(b), as we increase the white noise, IPC can still certify the entanglement dimensionality vector when the FBC fails with experimental fidelity $F<0.75$. The curves are induced from the noise model using experimental data; see Sec.~S2\,F of the Supplemental Material for details.
This highlights that QOT, by supplying all two-body marginals from only 25 settings, allows the IPC to augment conventional fidelity-based multipartite high-dimensional entanglement certification with improved noise resilience.

\textit{Discussion.—}In summary, we reported the implementation of high-dimensional overlapping tomography optimized by a graph-theoretical formulation. We experimentally demonstrated it on a four-photon entangled quantum state with local dimensions of (4,4,2,2). 
By reconstructing all two-body marginals with only 25 projective measurement settings, we quantified the entanglement properties of all two-body subsystems, providing a practical characterization method for applications such as layered quantum communication networks. 
Beyond the reduction in settings, we showed that these marginals, combined with the global fidelity, certify Schmidt numbers with higher noise tolerance than the fidelity alone. This route extends to $n$ parties, for which QOT provides all $k$-body marginals from $(d+1)^k\mathcal{O}(\log n)$ settings, so that the IPC can be evaluated for every bipartition in which one side contains at most $k$ parties.

QOT complements classical shadows \cite{huang2020predicting}, which also reconstruct the marginals of arbitrary states. Classical shadows choose the local bases at random for every copy, so with the same local MUBs they sample from 225 possible global settings. QOT instead uses 25 fixed settings, the minimum for local projective measurements. Applying the same confidence-region analysis to both schemes shows that QOT has the same sample-complexity scaling as classical shadows (Supplemental Material, Sec.~S1\,D). The reduction in settings is experimentally relevant because changing settings can take longer and introduce more setting errors than collecting further samples in a fixed configuration \cite{PRXQuantum.3.040310, wang2018quantum} (see also Supplemental Material, Sec.~S1\,E).

With continued advances in multidimensional quantum information processing \cite{kues2017chip, wang2018multidimensional, reimer2019high, luo2019quantum, cabrejo2023high}, our protocol offers an efficient and scalable framework, in terms of the number of measurement settings and the sample complexity, for subsystem reconstruction and global property learning in future complex quantum systems.
Our scheme can also be readily extended to quantum process overlapping tomography \cite{jkf7-wfcn} for qudit systems, with potential applications in benchmarking quantum gates and noise processes in quantum computing devices \cite{PhysRevApplied.18.064056, PhysRevLett.130.200601}.

\textit{Acknowledgments.—}We thank Satoya Imai and Zhen-Peng Xu for helpful discussions. This work is supported by ASTAR (M24M8b0004 and S24Q2d0009), Singapore National Research Foundation (NRF-CRP30-2023-0003, NRF-CRP31-0001, NRF2023-ITC004-001 and NRF-MSG-2023-0002), Singapore Ministry of Education Tier 2 Grant (MOE-T2EP50222-0018), Deutsche Forschungsgemeinschaft (DFG, German Research Foundation, project number 563437167), the Sino-German Center for Research Promotion (Project M-0294), and the German Federal Ministry of Research, Technology and Space (Projects QuKuK, Grant 
No.\ 16KIS1618K, BeRyQC, Grant No.\ 01N124023D, and PraQTisch, Grant No.\ 01N126223D).

\noindent\textit{Author contributions.—}
H. Wang and R. Qu contributed equally to this work.

\bibliography{bib-main} 

\end{document}


\title{Supplemental Material for: Experimental High-Dimensional Quantum Overlapping Tomography}

\author{Haifei Wang}
\affiliation{%
Centre for Quantum Technologies, Singapore 117543, Singapore}
\affiliation{%
Quantum Science and Engineering Centre (QSec), Nanyang Technological University, Singapore 639798, Singapore
}%

\author{Rui Qu}
\email{rui.qu@ntu.edu.sg}
\affiliation{%
Centre for Quantum Technologies, Singapore 117543, Singapore}
\affiliation{%
School of Electrical and Electronic Engineering, Nanyang Technological University, Singapore 639798, Singapore
}%

\author{Zhengning Yang}
\affiliation{%
Division of Physics and Applied Physics, School of Physical and Mathematical Sciences, Nanyang Technological University, Singapore, 637371, Singapore}

\author{Xiaodan Lyu}
\affiliation{%
School of Electrical and Electronic Engineering, Nanyang Technological University, Singapore 639798, Singapore
}%
\author{Haotao Zhu}
\affiliation{%
School of Electrical and Electronic Engineering, Nanyang Technological University, Singapore 639798, Singapore
}%

\author{Zitong Xu}
\affiliation{%
School of Electrical and Electronic Engineering, Nanyang Technological University, Singapore 639798, Singapore
}%

\author{Lianzhen Cao}
\affiliation{School of Physics and Electronic Information, Weifang University, Weifang 261061, China}

\author{Joel Yang}
\affiliation{Engineering Product Development, Singapore University of Technology and Design, Singapore 487372, Singapore}

\author{Otfried G\"uhne}
\email{otfried.guehne@uni-siegen.de}
\affiliation{Naturwissenschaftlich-Technische Fakult\"at, Universit\"at Siegen, Walter-Flex-Stra{\ss}e 3, 57068 Siegen, Germany}

\author{Weibo Gao}
\email{wbgao@ntu.edu.sg}
\affiliation{%
Centre for Quantum Technologies, Singapore 117543, Singapore}
\affiliation{%
Quantum Science and Engineering Centre (QSec), Nanyang Technological University, Singapore 639798, Singapore
}%
\affiliation{%
School of Electrical and Electronic Engineering, Nanyang Technological University, Singapore 639798, Singapore
}%
\affiliation{%
Division of Physics and Applied Physics, School of Physical and Mathematical Sciences, Nanyang Technological University, Singapore, 637371, Singapore}

\date{\today}

\maketitle  

\section{Protocol of High-dimensional Quantum Overlapping Tomography}

In this section, we describe the high-dimensional quantum overlapping tomography (QOT) protocol used in the main text and prove its scaling with the number of parties. 
We also discuss the relation between the number of samples and the confidence region.
Finally, we compare QOT with classical shadows and discuss the acquisition time, the experimental control complexity, and the classical reconstruction cost.

\subsection{Optimal settings and covering arrays}

We first formulate high-dimensional QOT as a covering-array problem \cite{colbourn2004combinatorial}.
Consider an $n$-partite system with local dimension $d$, where $d$ is a prime power so that each party admits a complete set of $(d+1)$ mutually unbiased bases (MUBs).
For party $i$, we label these bases by $\mathcal{B}^{(i)}_0,\mathcal{B}^{(i)}_1,\ldots,\mathcal{B}^{(i)}_d$.
Here $\mathcal{B}^{(i)}_x=\{\ket{b^{(i)}_{a|x}}\}_{a=0}^{d-1}$ denotes the $x$-th projective measurement basis, with basis label $x\in\{0,\ldots,d\}$ and outcome label $a\in\{0,\ldots,d-1\}$.

A global measurement setting is specified by choosing one local MUB label for every party.
The $\alpha$-th global setting is labeled as
\begin{equation}
    \mathbf{x}^{(\alpha)}
    =
    \left(
        x^{(\alpha)}_1,
        x^{(\alpha)}_2,
        \ldots,
        x^{(\alpha)}_n
    \right),
    \qquad
    x^{(\alpha)}_i\in\{0,\ldots,d\}.
\end{equation}
Thus, a measurement design with $N$ global settings can be represented by an $N\times n$ array
\begin{equation}
    A=\begin{pmatrix}
        x^{(1)}_1 & x^{(1)}_2 & \cdots & x^{(1)}_n \\
        x^{(2)}_1 & x^{(2)}_2 & \cdots & x^{(2)}_n \\
        \vdots    & \vdots    &        & \vdots    \\
        x^{(N)}_1 & x^{(N)}_2 & \cdots & x^{(N)}_n
    \end{pmatrix},
    \qquad
    x^{(\alpha)}_i\in\{0,\ldots,d\}.
\end{equation}
Rows label global measurement settings, while columns label parties.

Let $S=\{i_1,\ldots,i_k\}\subseteq\{1,2,...,n\}$ label a $k$-body subsystem of the $n$-body system.
Selecting the columns indexed by $S$ gives an $N\times k$ subarray which describes the MUB settings that are implemented on the marginal $\rho_S$.
The marginal $\rho_S$ can be reconstructed whenever the corresponding $N\times k$ subarray contains every tuple in $\{0,\ldots,d\}^k$ at least once. 
The problem of finding the smallest set of global measurement settings is then equivalent to the covering-array problem we introduce as follows.

A covering array $\mathrm{CA}(N;k,n,v)$ is an $N\times n$ array over an alphabet of size $v$ such that every $N\times k$ subarray contains all $v^k$ possible $k$-tuples at least once \cite{colbourn2004combinatorial}.
In the present MUB construction, the alphabet size is $v=d+1$. Therefore, the minimum number of settings is
\begin{equation}
    \Phi^{\mathrm{MUB}}_k (n,d)
    =
    \mathrm{CAN}(k,n,d+1),
    \label{eq:MUB_CAN}
\end{equation}
where $\mathrm{CAN}(k,n,v)$ denotes the minimum number of rows of a covering array $\mathrm{CA}(N;k,n,v)$.

This formulation is the MUB analogue of the covering-array formulation used in optimal Pauli \cite{hansenne2024optimal} and Gell-Mann-matrix overlapping tomography \cite{ma2026optimal}. 
The difference lies only in the local alphabet. 
For Pauli measurements on qubits, $v=3$. 
For the Gell-Mann-matrix construction of Ref.~\cite{ma2026optimal}, $v=d^2-1$. 
For the present MUB-based construction, each local setting is a $d$-outcome projective basis measurement, and hence $v=d+1$.

For $k=2$, Eq.~\eqref{eq:MUB_CAN} has an equivalent graph-theoretic formulation which is introduced in the main text. 
For systems with unequal local dimensions $d_1,\ldots,d_n$, one obtains a mixed covering array \cite{mixedca}. 
The alphabet size of column $i$ is then $v_i=d_i+1$. 
In the experiment reported in the main text, the local dimensions are $(4,4,2,2)$, so the corresponding alphabet sizes are $(5,5,3,3)$. The measurement settings used in the experiment are listed in Tab.~\ref{tab:settings}.

\begin{table}[ht]
\centering
\caption{The 25 measurement settings of two-body QOT of the $(4,4,2,2)$ state.}
\label{tab:settings}
\resizebox{\textwidth}{!}{%
\begin{tabular}{c*{25}{c}}
\toprule
Setting
 & 1 & 2 & 3 & 4 & 5 & 6 & 7 & 8 & 9 & 10
 & 11 & 12 & 13 & 14 & 15 & 16 & 17 & 18 & 19 & 20
 & 21 & 22 & 23 & 24 & 25 \\
\midrule
Qudit 1
& $\M_0$ & $\M_0$ & $\M_0$ & $\M_0$ & $\M_0$
& $\M_1$ & $\M_1$ & $\M_1$ & $\M_1$ & $\M_1$
& $\M_2$ & $\M_2$ & $\M_2$ & $\M_2$ & $\M_2$
& $\M_3$ & $\M_3$ & $\M_3$ & $\M_3$ & $\M_3$
& $\M_4$ & $\M_4$ & $\M_4$ & $\M_4$ & $\M_4$ \\

Qudit 2
& $\M_0$ & $\M_1$ & $\M_2$ & $\M_3$ & $\M_4$
& $\M_0$ & $\M_1$ & $\M_2$ & $\M_3$ & $\M_4$
& $\M_0$ & $\M_1$ & $\M_2$ & $\M_3$ & $\M_4$
& $\M_0$ & $\M_1$ & $\M_2$ & $\M_3$ & $\M_4$
& $\M_0$ & $\M_1$ & $\M_2$ & $\M_3$ & $\M_4$ \\

Qubit 1
& $X$ & $Y$ & $Y$ & $Z$ & $X$
& $Y$ & $Z$ & $X$ & $Y$ & $X$
& $X$ & $X$ & $Z$ & $Y$ & $Y$
& $Z$ & $Y$ & $X$ & $X$ & $Y$
& $Y$ & $X$ & $Y$ & $X$ & $Z$ \\

Qubit 2
& $Y$ & $Z$ & $X$ & $Y$ & $X$
& $Y$ & $X$ & $Y$ & $X$ & $Z$
& $X$ & $X$ & $Z$ & $Y$ & $Y$
& $X$ & $Y$ & $Y$ & $Z$ & $X$
& $Z$ & $Y$ & $X$ & $X$ & $Y$ \\
\bottomrule
\end{tabular}%
}
\end{table}

\subsection{Scaling of measurement settings}
\label{sec:scaling_settings}

We now prove the scaling $(d+1)^k \mathcal{O}(\log n)$ of the MUB-based construction. 
Because Eq.~\eqref{eq:MUB_CAN} identifies high-dimensional QOT with a covering-array problem of alphabet size $v=d+1$, the scaling follows from standard upper bounds on covering arrays, including the Stein--Lovász--Johnson bound \cite{sarkar2017upper}. 
For completeness, we give an elementary probabilistic estimation in the following.

Let $v=d+1$. Consider an $N\times n$ random array $X$, where each entry is chosen independently and uniformly from the alphabet $\{0,\ldots,v-1\}$. 
Fix a subset $S$ of $k$ columns and a fixed tuple $\mathbf{x}\in\{0,\ldots,v-1\}^k$.
For one row of the array, the probability that the entries in the columns $S$ equal $\mathbf{x}$ is $v^{-k}$. 
Therefore, the probability that $\mathbf{x}$ does not appear in any of the $N$ rows is
\begin{equation}
    \left(1-v^{-k}\right)^N
    \leq
    \exp\left(-\frac{N}{v^k}\right).
\end{equation}
There are $\binom{n}{k}$ choices of $S$ and $v^k$ possible tuples for each choice of $S$. 
Using the union bound, the probability that the random array fails to be a covering array is bounded by
\begin{equation}
    p_{\mathrm{fail}}
    \leq
    \binom{n}{k}v^k
    \exp\left(-\frac{N}{v^k}\right).
\end{equation}
Requiring $p_{\mathrm{fail}} \le \delta$, it suffices to choose
\begin{equation}
    N
    \ge
    v^k
    \left[
        \log\binom{n}{k}
        +
        k\log v
        +
        \log\left(\frac{1}{\delta}\right)
    \right],
\end{equation}
Therefore, an $N\times n$ covering array exists with high probability $(1-\delta)$. 
Substituting $v=d+1$ gives the minimum number of settings required
\begin{equation}
    \Phi^{\mathrm{MUB}}_k(n,d)
    =
    (d+1)^k
    \left[
        \log\binom{n}{k}
        +
        k\log(d+1)
        +
        \log\left(\frac{1}{\delta}\right)
    \right].
\end{equation}
For fixed $k$, this gives
\begin{equation}
    \Phi^{\mathrm{MUB}}_k(n,d)
    =
    (d+1)^k \mathcal{O}(\log n).
    \label{eq:MUB_scaling}
\end{equation}

A sharper form follows directly from the Stein--Lovász--Johnson \cite{sarkar2017upper} bound for covering arrays, which gives
\begin{equation}
    \mathrm{CAN}(k,n,v)
    \leq
    \frac{k}{\log\!\left(\frac{v^k}{v^k-1}\right)}
    \log n \, [1+o(1)].
    \label{eq:SLJ_MUB}
\end{equation}
Since
\begin{equation}
    \log\!\left(\frac{v^k}{v^k-1}\right)
    =
    \log\!\left(1+\frac{1}{v^k-1}\right) \sim \frac{1}{v^k-1} \qquad \left(v^k \gg 1\right),
\end{equation}
the prefactor in Eq.~\eqref{eq:SLJ_MUB} is of order $v^k=(d+1)^k$, which again gives Eq.~\eqref{eq:MUB_scaling}.

This should be compared with independent marginal tomography. 
Each $k$-qudit marginal requires $(d+1)^k$ tensor-product MUB settings, and there are $\binom{n}{k}$ such marginals. 
Thus, the naive number of settings is $(d+1)^k \binom{n}{k}$.
The QOT construction reduces this to $(d+1)^k\mathcal{O}(\log n)$, changing the dependence on the number of parties $n$ from polynomial to logarithmic.

\subsection{Confidence region and sample complexity}
\label{sec:confidence_sample}

In this part, we discuss how many copies of the state are required to reconstruct the desired marginals within a prescribed statistical accuracy.
This is a different question from the one addressed above, which asks how many measurement settings are needed to reconstruct every marginal.

We use the confidence-region method of Ref.~\cite{PhysRevA.109.062417}, which was also used in the Supplemental Material of Ref.~\cite{hansenne2024optimal} to compare overlapping tomography schemes.
For each marginal $S$, the observed data give an estimate $\hat{\rho}_S$ of the unknown marginal state $\rho_S$.
Since the data have statistical fluctuations, $\hat{\rho}_S$ fluctuates around the true state $\rho_S$.

Ref.~\cite{PhysRevA.109.062417} gives a confidence region of the form
\begin{equation}
    \Pr\!\left[
        \left\|\hat{\rho}_S-\rho_S\right\|_2
        \leq \tau\,\sigma_S
    \right]\geq 1-\delta .
    \label{eq:confidence_region}
\end{equation}
Here $\|\cdot\|_2$ is the Hilbert--Schmidt norm, $\delta$ is the allowed failure probability, and $\tau=3\sqrt{u}(\sqrt{u}+\sqrt{u+1})$ with $u=2\log(8/\delta)/(9N)$. 
The constant $\sigma_S$ depends only on the measurement settings. 
Specifically, let $E_{a|s}=q_s\Pi_{a|s}$ be the generalized measurement effect obtained from the projector $\Pi_{a|s}$ measured in setting $s$ sampled with relative frequency $q_s$, with $\sum_s q_s = 1$. Then $\{E_{a|s}\}$ forms a single informationally complete POVM. Define the measurement map by $(\mathcal{M}_S \varrho)_{a,s}=\operatorname{tr}(E_{a|s}\varrho)$, which sends a state $\varrho$ to the outcome probabilities labeled by $(a,s)$. Its pseudoinverse $\mathcal{M}_S^{+}$ reconstructs the state from these probabilities and thus has columns indexed by $(a,s)$ and rows indexed by the components of the vectorized state. Then $\sigma_S$ is the largest Hilbert--Schmidt norm of the operators represented by the columns of the pseudoinverse $\mathcal{M}_S^{+}$ (see Ref.~\cite{PhysRevA.109.062417} for details). This confidence region assumes that the measured copies are independent and identically distributed, in particular that the source is stationary during the acquisition, and it does not cover variations of the state between trials. The same assumption underlies the Poissonian Monte Carlo uncertainties reported in the main text and in Sec.~S2.

From Eq.~\eqref{eq:confidence_region}, requiring $\|\hat{\rho}_S-\rho_S\|_2\leq\epsilon$ for one marginal gives $N_S=O[(\sigma_S^2/\epsilon^2)\log(1/\delta)]$.
Using a union bound over all marginals gives the sample complexity for QOT
\begin{equation}
    N
    =
    O\!\left(
        \frac{\sigma_{\max}^2}{\epsilon^2}
        \log\frac{|K|}{\delta}
    \right),
    \label{eq:sample_complexity}
\end{equation}
where $K$ is the set of marginals of interest, and $\sigma_{\max}:=\max_{S\in K}\sigma_S$ is the worst case of $\sigma_S$. The measurement settings used in the experiment (Tab.~\ref{tab:settings}) have $\sigma_{\max}=19$. For context, the standard 9 two-body Pauli settings for two-qubit tomography have $\sigma_{\text{Pauli}} = 5$.

For a complete set of MUB measurements on a subsystem of Hilbert-space dimension $D_S$, we have $\sigma_S=O(D_S)$ \cite{PhysRevA.109.062417}.
In the homogeneous case $d_i=d$, this means $\sigma_{\max}=O(d^k)$, while for all $k$-body marginals $|K|=\binom{n}{k}$.
Substituting these into Eq.~\eqref{eq:sample_complexity} gives the sample complexity
\begin{equation}
    N
    =
    O\!\left[
        \frac{d^{2k}}{\epsilon^2}
        \log\!\left(\frac{\binom{n}{k}}{\delta}\right)
    \right]
    =
    O\left[\frac{d^{2k}}{\epsilon^2}\left(k\log n+\log\frac{1}{\delta}\right)\right]
    .
\end{equation}
Thus, for fixed local dimension $d$ and fixed marginal size $k$, the number of samples grows only logarithmically with the total number of parties $n$.

\subsection{Comparison with classical shadows}
\label{sec:comparison_shadows}

QOT and classical shadows can play complementary roles in efficient marginal reconstruction. Here we compare their statistical costs under the same local measurement constraints, using the complete local MUBs available in our experiment. In the classical-shadow protocol, each party independently chooses a basis with uniform probability for every sample. In QOT, equal numbers of samples are assigned to the 25 global settings in Table~\ref{tab:settings}. Both protocols use independent copies of the same arbitrary state. Within each protocol, every global measurement record contributes to reconstructing all six two-body marginals. The target accuracy and joint confidence level are also the same. This comparison concerns the sufficient sample count for linear reconstruction given by the confidence region below, not the errors of the maximum-likelihood reconstruction used in the experiment.

To make this comparison, we apply the confidence-region method of Ref.~\cite{PhysRevA.109.062417}, already used in Sec.~\ref{sec:confidence_sample}, to both reconstruction schemes. For classical shadows, the reconstruction itself follows Refs.~\cite{huang2020predicting,wang2023classicalmub}, in which each recorded basis and outcome is converted into a matrix and the sample average estimates the state. Under uniform MUB sampling, the reconstruction rule in Eqs. (2)--(3) of Ref.~\cite{wang2023classicalmub} gives a local matrix with eigenvalue \(d_i\) along the observed basis vector and eigenvalue \(-1\) along each of the remaining \(d_i-1\) directions. Its squared Hilbert--Schmidt norm is therefore \(d_i^2+d_i-1\). Since the reconstruction of a marginal uses the tensor product of these local matrices, their norms multiply. For a marginal \(S\), the resulting confidence-region parameter is

\begin{equation}
\sigma_S
=\left[\prod_{i\in S}(d_i^2+d_i-1)\right]^{1/2}.
\label{eq:shadow_sigma_comparison}
\end{equation}

For \(k\)-body marginals with local dimension \(d\), this gives \(\sigma_S=(d^2+d-1)^{k/2}=\mathcal{O}(d^k)\) for fixed \(k\), which matches the scaling for QOT in Sec.~\ref{sec:confidence_sample}. Using the same confidence bound and assigning failure probability \(\delta/|K|\) to each marginal, both schemes therefore require \(\mathcal{O}\bigl(d^{2k}(k\log n+\log(1/\delta))/\epsilon^{2}\bigr)\) samples for a prescribed Hilbert--Schmidt error and joint confidence level. Thus, within this common confidence-region analysis, the optimized QOT schedule preserves the sample-complexity scaling of the classical-shadow reconstruction.

With the same sample-complexity scaling, the advantage of QOT lies in the number of measurement settings. QOT uses 25 fixed global settings, whereas the random local basis choices in the classical-shadow protocol generate \(5\times5\times3\times3=225\) possible global combinations. The 25 settings are the minimum for any scheme based on local projective measurements. For a two-body marginal with local dimensions \(d_i\) and \(d_j\), each setting determines at most \((d_i-1)(d_j-1)\) of its \((d_i^2-1)(d_j^2-1)\) correlation parameters, since a local basis with \(d\) outcomes gives only \(d-1\) independent probabilities. At least \((d_i+1)(d_j+1)\) settings are therefore required, that is, 25 for \(\rho_{AB}\). The same counting gives at least \(\prod_{i\in S}(d_i+1)\) settings for a \(k\)-body marginal \(S\).

These results show that QOT has the same sample-complexity scaling as classical shadows while using the minimum number of settings for local projective measurements. As discussed in the main text, changing measurement settings can be more time-consuming and error-prone than repeating measurements in a fixed configuration. The setting reduction is therefore relevant not only to acquisition time but also to the control of experimental errors.

\subsection{Acquisition time, control complexity, and reconstruction cost}
\label{sec:other_resources}

The acquisition time mentioned above is the total duration of the measurements. It consists of the integration time and the total setting-switching time. At a given event rate, the integration time is proportional to the number of samples. If all switches have the same duration, the total switching time is proportional to the number of settings. Each party changes only its own local basis, so the parties can switch in parallel and the duration of each switch does not grow with \(n\). For fixed \(d\), \(k\), \(\epsilon\), and \(\delta\), the number of samples (Sec.~\ref{sec:confidence_sample}) and the number of settings (Sec.~\ref{sec:scaling_settings}) grow only logarithmically with \(n\), so at a given event rate the acquisition time also grows only logarithmically. The event rate may decrease with \(n\), as in multiphoton experiments. Such a decrease is set by the source, not by the tomography scheme, and lengthens the integration time of any scheme by the same factor.

As for the experimental control complexity, changing a setting in our experiment means reconfiguring the wave plates of each party to one of its fixed local bases (see Table~\ref{tab:waveplate-settings} for the ququarts). Each party thus needs only its \(d_i+1\) local bases, independent of \(n\), and for local dimension \(d\) the number of global settings grows as \((d+1)^k\mathcal{O}(\log n)\). In this sense, the control complexity also grows only logarithmically with \(n\).

QOT is designed to reduce the measurement resources discussed above, not the classical reconstruction cost. Even so, this cost grows only polynomially with \(n\) for fixed \(d\) and \(k\), rather than exponentially as for full-state tomography. This is because each marginal is reconstructed separately by maximum-likelihood estimation (Sec.~\ref{sec:mle_marginals}), and for \(n\) parties there are \(\binom{n}{k}=\mathcal{O}(n^k)\) marginals, each of dimension \(d^k\).

\section{Detailed experimental methods and results}

\subsection{Generation of the target state}

\begin{figure}[h]
    \centering
    \includegraphics[width=.3\linewidth]{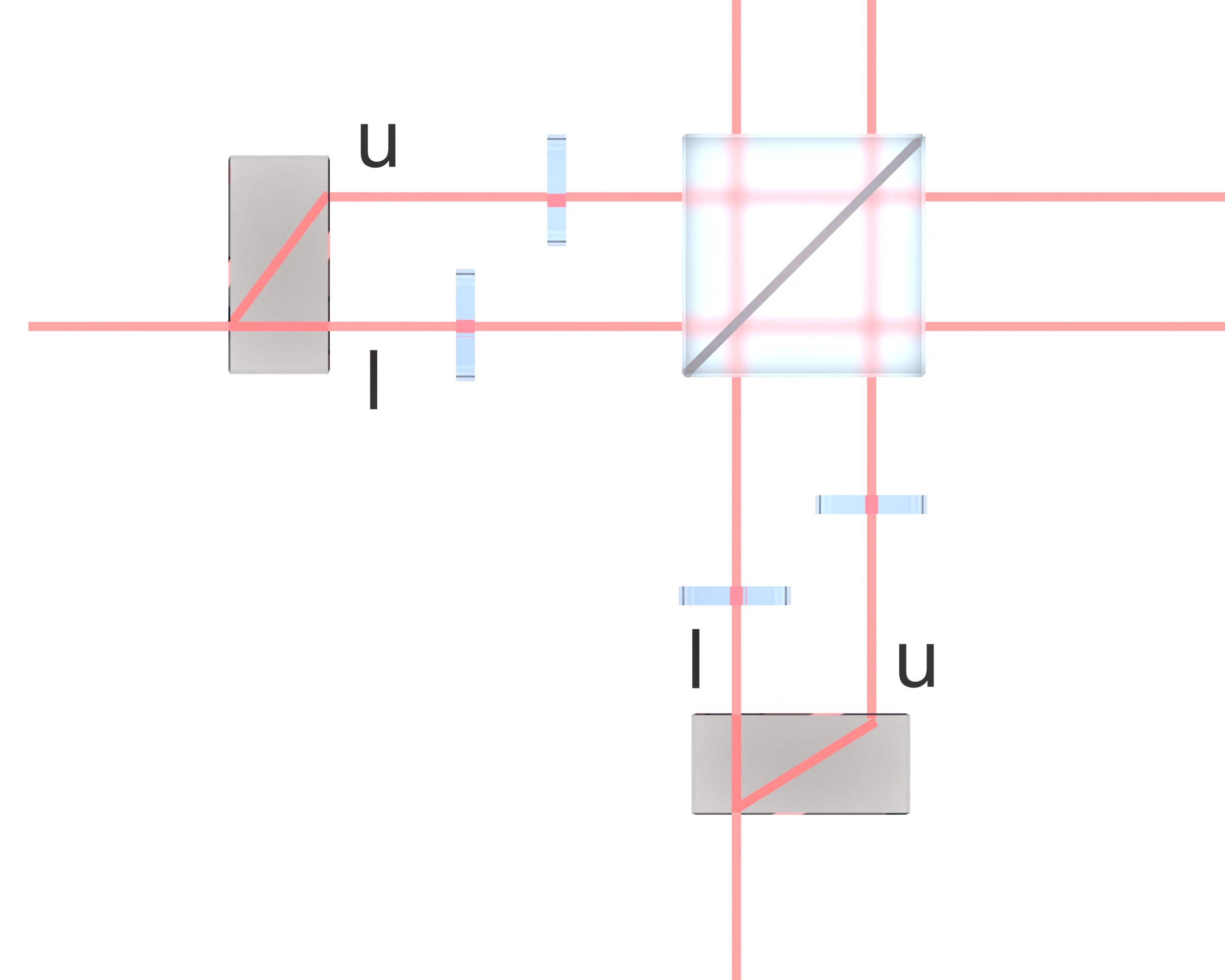}
    \caption{Path-polarization dimension-increasing device.}
    \label{fig:SM_fig_di.jpg}
\end{figure}

In this section, we describe the generation of the target state using the path-polarization dimension-increasing device shown in Fig.~\ref{fig:SM_fig_di.jpg}. It consists of beam displacers (BDs), half-wave plates (HWP), and a polarization beam splitter (PBS). The BDs displace vertically polarized photons into the upper path while leaving horizontally polarized photons in the lower path. If we feed a Bell pair $(|HH\rangle + |VV\rangle)/\sqrt{2}$ into both arms of the device, then after the BDs it becomes $(|H_lH_l\rangle + |V_uV_u\rangle)/\sqrt{2}$. With all HWPs placed at $22.5^\circ$, the state evolves to 
\begin{equation}
    \frac{1}{\sqrt{2}} \left(\frac{|H_l\rangle + |V_l\rangle}{\sqrt{2}}\frac{|H_l\rangle + |V_l\rangle}{\sqrt{2}} + \frac{|H_u\rangle - |V_u\rangle}{\sqrt{2}}\frac{|H_u\rangle - |V_u\rangle}{\sqrt{2}}\right).
\end{equation}

After the PBS, we select events where there is one photon on both output arms. This postselection prepares the state
\begin{equation}
    \frac{1}{2}(|H_lH_l\rangle + |V_lV_l\rangle + |H_uH_u\rangle + |V_uV_u\rangle).
\end{equation}

Similarly, if we feed two photons from the 4-photon GHZ state $(|HHHH\rangle + |VVVV\rangle)/\sqrt{2}$ into this dimension-increasing device, then we prepare the target state
\begin{equation}
    \frac{1}{2}(|H_lH_lHH\rangle + |V_lV_lHH\rangle + |H_uH_uVV\rangle + |V_uV_uVV\rangle).
\end{equation}

With the label $|H_u\rangle \rightarrow |0\rangle$, $|H_l\rangle \rightarrow |1\rangle$, $|V_u\rangle \rightarrow |2\rangle$, and $|V_l\rangle \rightarrow |3\rangle$, the state is encoded as
\begin{equation}
    |\Lambda\rangle =\frac{1}{2}(|1100\rangle + |3300\rangle + |0011\rangle + |2211\rangle).
    \label{eq:Psi}
\end{equation}

\subsection{Fidelity measurement}

In this section, we describe how we measure the fidelity of the experimentally generated state $\rho$ with respect to the target state
\begin{equation}
    |\Lambda\rangle
    =
    \frac{1}{2}
    \left(
        |1100\rangle
        +
        |3300\rangle
        +
        |0011\rangle
        +
        |2211\rangle
    \right).
\end{equation}

Since the target state is pure, the fidelity is $F(\rho,|\Lambda\rangle\langle\Lambda|)=\langle\Lambda|\rho|\Lambda\rangle$. 
Expanding in the computational basis gives
\begin{equation}
    F(\rho,|\Lambda\rangle\langle\Lambda|)
    =
    \frac{1}{4}
    \sum_{\mu,\nu\in\mathcal{I}}
    \langle \mu|\rho|\nu\rangle ,
\end{equation}
where $\mathcal{I}=\{1100,\,3300,\,0011,\,2211\}$.
Thus, the fidelity is determined by four diagonal populations and the real parts of the six independent coherences:
\begin{equation}
    F(\rho,|\Lambda\rangle\langle\Lambda|)
    =
    \frac{1}{4}
    \left[
        \sum_{\mu\in\mathcal{I}}
        \rho_{\mu,\mu}
        +
        2
        \sum_{\substack{\mu,\nu\in\mathcal{I}\\ \mu<\nu}}
        \mathrm{Re}(\rho_{\mu,\nu})
    \right].
    \label{eq:fidelity}
\end{equation}

\begin{figure}[h]
    \centering
    \includegraphics[width=.8\linewidth]{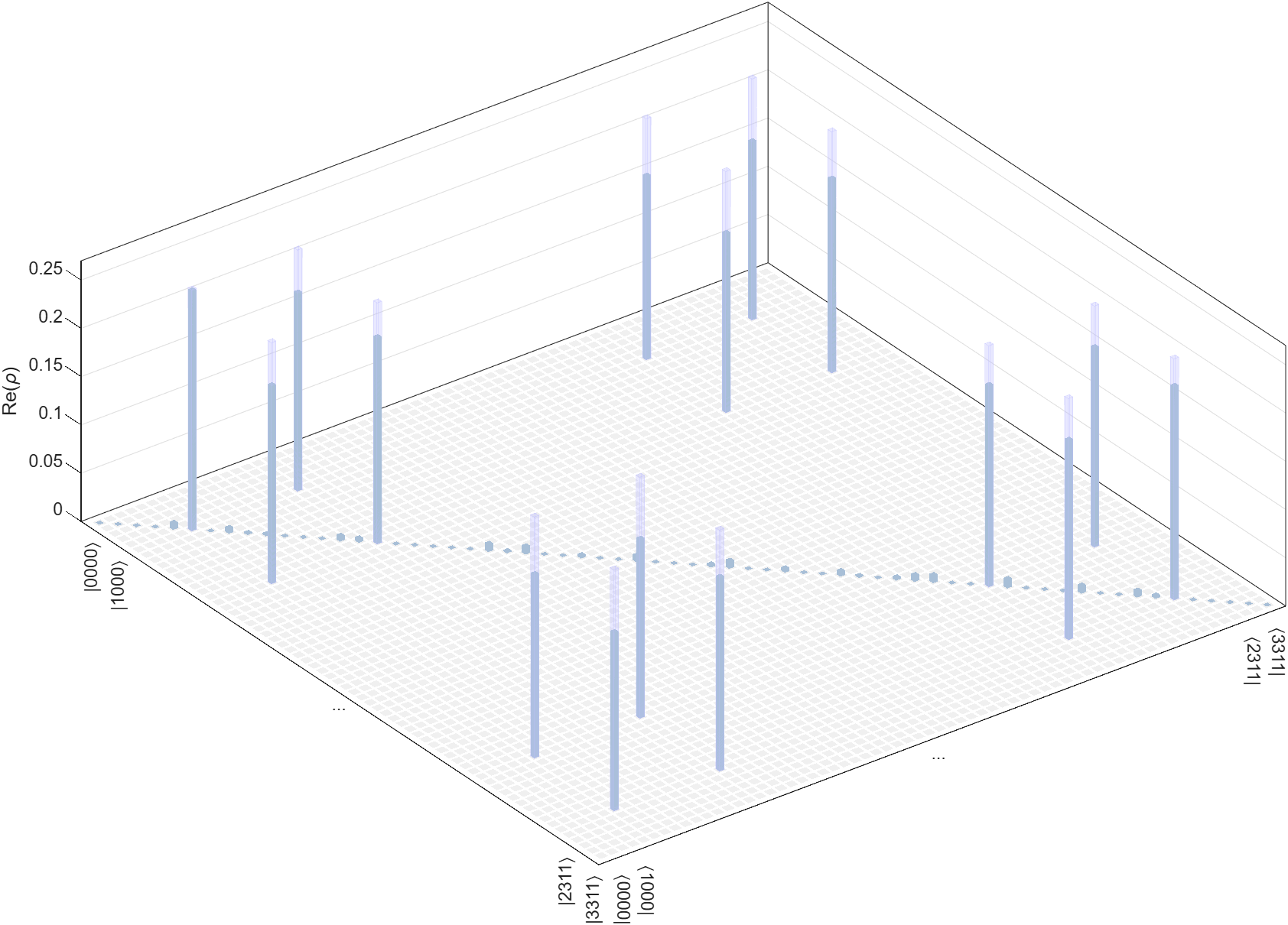}
    \caption{The 64 diagonal elements and 6 unique real parts of off-diagonal elements of experimental state $\rho$. These elements determine the global fidelity $F=0.809\pm 0.009$. The remaining elements are not measured and left blank.}
    \label{fig:SM_fig_fidelity.jpg}
\end{figure}

\begin{table*}[ht]
\centering
\caption{Measured density matrix elements of the experimental state
(real part). The bold elements have overlap with the target state and can be used to calculate the fidelity.}
\scriptsize
\renewcommand{\arraystretch}{1.15}
\setlength{\tabcolsep}{3pt}
\resizebox{\textwidth}{!}{%
\begin{tabular}{|c|c|c|c|c|c|c|c|}
\hline
$|0000\rangle\langle0000|$ & $|1000\rangle\langle1000|$ & $|2000\rangle\langle2000|$ & $|3000\rangle\langle3000|$ & $|0100\rangle\langle0100|$ & $\mathbf{|1100\rangle\langle1100|}$ & $|2100\rangle\langle2100|$ & $|3100\rangle\langle3100|$ \\
\hline
$0.0000 \pm 0.0000$ & $0.0000 \pm 0.0000$ & $0.0000 \pm 0.0000$ & $0.0000 \pm 0.0000$ & $0.0062 \pm 0.0028$ & $\mathbf{0.2484 \pm 0.0153}$ & $0.0000 \pm 0.0000$ & $0.0050 \pm 0.0025$ \\
\hline
$|0200\rangle\langle0200|$ & $|1200\rangle\langle1200|$ & $|2200\rangle\langle2200|$ & $|3200\rangle\langle3200|$ & $|0300\rangle\langle0300|$ & $|1300\rangle\langle1300|$ & $|2300\rangle\langle2300|$ & $\mathbf{|3300\rangle\langle3300|}$ \\
\hline
$0.0012 \pm 0.0012$ & $0.0012 \pm 0.0012$ & $0.0000 \pm 0.0000$ & $0.0000 \pm 0.0000$ & $0.0000 \pm 0.0000$ & $0.0050 \pm 0.0025$ & $0.0037 \pm 0.0022$ & $\mathbf{0.2135 \pm 0.0145}$ \\
\hline
$|0010\rangle\langle0010|$ & $|1010\rangle\langle1010|$ & $|2010\rangle\langle2010|$ & $|3010\rangle\langle3010|$ & $|0110\rangle\langle0110|$ & $|1110\rangle\langle1110|$ & $|2110\rangle\langle2110|$ & $|3110\rangle\langle3110|$ \\
\hline
$0.0000 \pm 0.0000$ & $0.0000 \pm 0.0000$ & $0.0000 \pm 0.0000$ & $0.0000 \pm 0.0000$ & $0.0000 \pm 0.0000$ & $0.0075 \pm 0.0030$ & $0.0012 \pm 0.0012$ & $0.0075 \pm 0.0030$ \\
\hline
$|0210\rangle\langle0210|$ & $|1210\rangle\langle1210|$ & $|2210\rangle\langle2210|$ & $|3210\rangle\langle3210|$ & $|0310\rangle\langle0310|$ & $|1310\rangle\langle1310|$ & $|2310\rangle\langle2310|$ & $|3310\rangle\langle3310|$ \\
\hline
$0.0000 \pm 0.0000$ & $0.0000 \pm 0.0000$ & $0.0025 \pm 0.0018$ & $0.0000 \pm 0.0000$ & $0.0000 \pm 0.0000$ & $0.0062 \pm 0.0028$ & $0.0000 \pm 0.0000$ & $0.0000 \pm 0.0000$ \\
\hline
$|0001\rangle\langle0001|$ & $|1001\rangle\langle1001|$ & $|2001\rangle\langle2001|$ & $|3001\rangle\langle3001|$ & $|0101\rangle\langle0101|$ & $|1101\rangle\langle1101|$ & $|2101\rangle\langle2101|$ & $|3101\rangle\langle3101|$ \\
\hline
$0.0000 \pm 0.0000$ & $0.0025 \pm 0.0018$ & $0.0075 \pm 0.0030$ & $0.0000 \pm 0.0000$ & $0.0000 \pm 0.0000$ & $0.0037 \pm 0.0022$ & $0.0000 \pm 0.0000$ & $0.0000 \pm 0.0000$ \\
\hline
$|0201\rangle\langle0201|$ & $|1201\rangle\langle1201|$ & $|2201\rangle\langle2201|$ & $|3201\rangle\langle3201|$ & $|0301\rangle\langle0301|$ & $|1301\rangle\langle1301|$ & $|2301\rangle\langle2301|$ & $|3301\rangle\langle3301|$ \\
\hline
$0.0050 \pm 0.0025$ & $0.0012 \pm 0.0012$ & $0.0000 \pm 0.0000$ & $0.0025 \pm 0.0018$ & $0.0062 \pm 0.0028$ & $0.0075 \pm 0.0030$ & $0.0000 \pm 0.0000$ & $0.0000 \pm 0.0000$ \\
\hline
$\mathbf{|0011\rangle\langle0011|}$ & $|1011\rangle\langle1011|$ & $|2011\rangle\langle2011|$ & $|3011\rangle\langle3011|$ & $|0111\rangle\langle0111|$ & $|1111\rangle\langle1111|$ & $|2111\rangle\langle2111|$ & $|3111\rangle\langle3111|$ \\
\hline
$\mathbf{0.2085 \pm 0.0144}$ & $0.0087 \pm 0.0033$ & $0.0000 \pm 0.0000$ & $0.0000 \pm 0.0000$ & $0.0000 \pm 0.0000$ & $0.0075 \pm 0.0030$ & $0.0000 \pm 0.0000$ & $0.0000 \pm 0.0000$ \\
\hline
$|0211\rangle\langle0211|$ & $|1211\rangle\langle1211|$ & $\mathbf{|2211\rangle\langle2211|}$ & $|3211\rangle\langle3211|$ & $|0311\rangle\langle0311|$ & $|1311\rangle\langle1311|$ & $|2311\rangle\langle2311|$ & $|3311\rangle\langle3311|$ \\
\hline
$0.0062 \pm 0.0028$ & $0.0025 \pm 0.0018$ & $\mathbf{0.2210 \pm 0.0147}$ & $0.0000 \pm 0.0000$ & $0.0000 \pm 0.0000$ & $0.0000 \pm 0.0000$ & $0.0000 \pm 0.0000$ & $0.0000 \pm 0.0000$ \\
\hline
$\mathbf{|1100\rangle\langle3300|}$ & $\mathbf{|1100\rangle\langle0011|}$ & $\mathbf{|1100\rangle\langle2211|}$ & $\mathbf{|3300\rangle\langle0011|}$ & $\mathbf{|3300\rangle\langle2211|}$ & $\mathbf{|0011\rangle\langle2211|}$ &  &  \\
\hline
$\mathbf{0.2050 \pm 0.0053}$ & $\mathbf{0.1900 \pm 0.0030}$ & $\mathbf{0.1842 \pm 0.0031}$ & $\mathbf{0.1853 \pm 0.0035}$ & $\mathbf{0.2005 \pm 0.0031}$ & $\mathbf{0.2064 \pm 0.0055}$ &  &  \\
\hline
\end{tabular}%
}
\label{tab:rho-real}
\end{table*}

The four populations $\rho_{\mu,\mu}$ are obtained by measuring all parties in the computational basis. 
To measure the coherences $\rho_{\mu,\nu}$, we use Pauli operators acting on the relevant subspaces. 
For $a\neq b$, define
\begin{equation}
    X_{a,b}
    =
    |a\rangle\langle b|
    +
    |b\rangle\langle a|,
    \qquad
    Y_{a,b}
    =
    -i|a\rangle\langle b|
    +
    i|b\rangle\langle a|.
\end{equation}
Then
\begin{equation}
    |a\rangle\langle b|
    =
    \frac{1}{2}
    \left(
        X_{a,b}
        +
        iY_{a,b}
    \right),
    \qquad
    |b\rangle\langle a|
    =
    \frac{1}{2}
    \left(
        X_{a,b}
        -
        iY_{a,b}
    \right).
\end{equation}

For a coherence between two four-party basis states
\begin{equation}
    |a_1a_2a_3a_4\rangle
    \quad\mathrm{and}\quad
    |b_1b_2b_3b_4\rangle ,
\end{equation}
where $a_i\neq b_i$ for all four parties, the real part is obtained by
\begin{equation}
\begin{aligned}
    &\langle a_1a_2a_3a_4|\rho|b_1b_2b_3b_4\rangle
    +
    \langle b_1b_2b_3b_4|\rho|a_1a_2a_3a_4\rangle
    \\
    &=
    \frac{1}{8}
    \Big(
        \langle X_{a_1,b_1}X_{a_2,b_2}X_{a_3,b_3}X_{a_4,b_4}\rangle
        -
        \langle X_{a_1,b_1}X_{a_2,b_2}Y_{a_3,b_3}Y_{a_4,b_4}\rangle
        \\
        &\qquad
        -
        \langle X_{a_1,b_1}Y_{a_2,b_2}X_{a_3,b_3}Y_{a_4,b_4}\rangle
        -
        \langle X_{a_1,b_1}Y_{a_2,b_2}Y_{a_3,b_3}X_{a_4,b_4}\rangle
        \\
        &\qquad
        -
        \langle Y_{a_1,b_1}X_{a_2,b_2}X_{a_3,b_3}Y_{a_4,b_4}\rangle
        -
        \langle Y_{a_1,b_1}X_{a_2,b_2}Y_{a_3,b_3}X_{a_4,b_4}\rangle
        \\
        &\qquad
        -
        \langle Y_{a_1,b_1}Y_{a_2,b_2}X_{a_3,b_3}X_{a_4,b_4}\rangle
        +
        \langle Y_{a_1,b_1}Y_{a_2,b_2}Y_{a_3,b_3}Y_{a_4,b_4}\rangle
    \Big).
    \label{eq:four_body_coherence}
\end{aligned}
\end{equation}
Here $\langle O\rangle=\mathrm{Tr}(\rho O)$. 
Equation~\eqref{eq:four_body_coherence} is used for measuring the off-diagonal elements
\begin{equation}
    \ket{1100}\bra{0011}, \quad 
    \ket{1100}\bra{2211}, \quad
    \ket{3300}\bra{0011}, \quad
    \ket{3300}\bra{2211}
\end{equation}
which differ on all four parties.

For the remaining off-diagonal elements
\begin{equation}
    \ket{1100}\bra{3300}, \quad \ket{0011}\bra{2211}
\end{equation}
only the two qudit parties change, while the qubit parties remain fixed in the computational basis. 
For these terms, we use the corresponding two-body expression conditioned on the computational-basis outcomes of the two qubits:
\begin{equation}
\begin{aligned}
    &\langle a_1a_2c_3c_4|\rho|b_1b_2c_3c_4\rangle
    +
    \langle b_1b_2c_3c_4|\rho|a_1a_2c_3c_4\rangle
    \\
    &=
    \frac{1}{2}
    \left[
        \left\langle
            X_{a_1,b_1}X_{a_2,b_2}
            \Pi_{c_3}
            \Pi_{c_4}
        \right\rangle
        -
        \left\langle
            Y_{a_1,b_1}Y_{a_2,b_2}
            \Pi_{c_3}
            \Pi_{c_4}
        \right\rangle
    \right],
    \label{eq:two_qudit_coherence}
\end{aligned}
\end{equation}
where $\Pi_{c_j}=|c_j\rangle\langle c_j|$
is the computational-basis projector on qubit party $j$.

In the experiment, we measure the four diagonal populations and the six unique coherence terms in Eq.~\eqref{eq:fidelity}, which are listed in Tab.~\ref{tab:rho-real} and shown in Fig.~\ref{fig:SM_fig_fidelity.jpg}.
The required observables are grouped into 35 measurement settings. 
For each setting, the integration time is 900 seconds. 
The measured populations and real parts of the coherences are then substituted into Eq.~\eqref{eq:fidelity}, yielding the global fidelity $0.809\pm 0.009$ reported in the main text.
The uncertainty is obtained by Monte Carlo resampling (1000 trials) of the photon-counting data assuming Poissonian statistics.

\subsection{Implementation of mutually unbiased basis measurements}

A set of orthonormal bases is called mutually unbiased bases (MUB) if $|\langle \psi | \phi \rangle|^2=1/d$ holds for any two vectors $|\psi\rangle$ and $|\phi\rangle$ belonging to two different bases. For a prime-power dimension $d$, a complete set of $(d+1)$ MUBs exists, which is informationally complete for quantum state tomography on a qudit \cite{lima2011experimental, PhysRevLett.110.143601}.

\begin{figure}[h]
    \centering
    \includegraphics[width=0.5\linewidth]{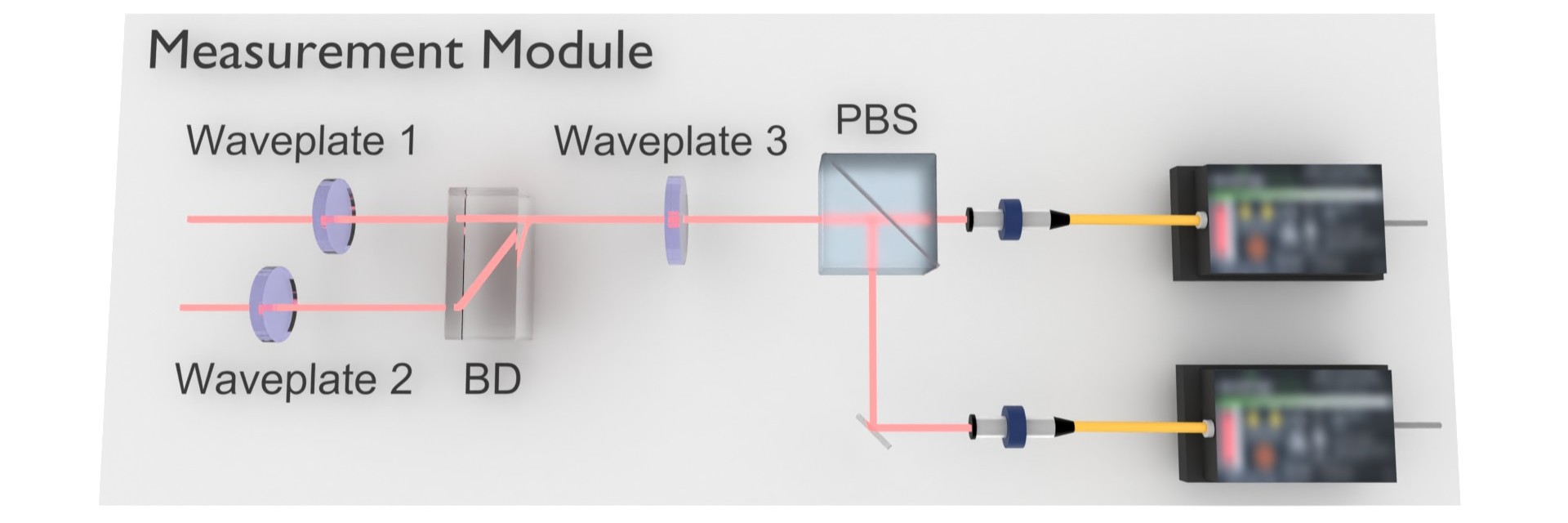}
    \caption{Measurement module for implementing the MUB measurements of the path-polarization hybrid qudit. BD: beam displacer, PBS: polarization beam splitter. The waveplate configuration is listed in Tab.~\ref{tab:waveplate-settings}.}
    \label{fig:SM_fig_MP.jpg}
\end{figure}

In this section, we detail the implementation of the MUB measurements for our path-polarization qudit with local dimension $d=4$. In this case, there are $d+1=5$ MUBs
\begin{equation}
\begin{aligned}
    \M_0 = \begin{pmatrix}
        1 & 0 & 0 & 0 \\
        0 & 1 & 0 & 0 \\
        0 & 0 & 1 & 0 \\
        0 & 0 & 0 & 1 
    \end{pmatrix}, \quad
    \M_1 = \frac{1}{2}\begin{pmatrix}
        1 & 1 & 1 & 1 \\
        1 & -1 & -1 & 1 \\
        1 & 1 & -1 & -1 \\
        1 & -1 & 1 & -1 
    \end{pmatrix}, \quad
    \M_2 = \frac{1}{2}\begin{pmatrix}
        1 & 1 & 1 & 1 \\
        -i & i & i & -i \\
        1 & 1 & -1 & -1 \\
        i & -i & i & -i 
    \end{pmatrix}, \\
    \M_3 = \frac{1}{2}\begin{pmatrix}
        1 & 1 & 1 & 1 \\
        i & -i & -i & i \\
        i & i & -i & -i \\
        -1 & 1 & -1 & 1 
    \end{pmatrix}, \quad
    \M_4 = \frac{1}{2}\begin{pmatrix}
        1 & 1 & 1 & 1 \\
        1 & -1 & -1 & 1 \\
        i & i & -i & -i \\
        -i & i & -i & i 
    \end{pmatrix},
\end{aligned}
\end{equation}
whose column vectors form an orthonormal basis.
These MUB measurements are implemented by the measurement module shown in Fig.~\ref{fig:SM_fig_MP.jpg}. 

We encode the hybrid path--polarization states as $|H_u\rangle \rightarrow |0\rangle$, $|H_l\rangle \rightarrow |1\rangle$, $|V_u\rangle \rightarrow |2\rangle$, and $|V_l\rangle \rightarrow |3\rangle$. 
Measurements in the computational basis $\M_0$ $(\ket{H_u},\ket{H_l},\ket{V_u},\ket{V_l})$ are performed as follows: if we place a half-wave plate (HWP) at $\pi / 4$ at the lower path before the BD, the two detectors then analyze $\ket{H_u}$ and $\ket{H_l}$, respectively. If we instead place the HWP at the upper path, then the two detectors analyze $\ket{V_u}$ and $\ket{V_l}$, respectively (see Tab.~\ref{tab:waveplate-settings}). Superpositions of computational-basis states are analyzed by applying HWP or quarter-wave plates (QWPs) to suitable spatial modes. In fact, all of the remaining MUBs $M_{1,2,3,4}$ can be realized by placing appropriate wave plates at the locations indicated in Fig.~\ref{fig:SM_fig_MP.jpg}, where the waveplates 1 and 2 interfere the polarization modes and the waveplate 3 interferes the path modes. The configuration of the waveplates for the MUBs $\{M_k\}_{k=0}^{d}$ are listed in Tab.~\ref{tab:waveplate-settings}.

\begin{table}[ht]
\centering
\caption{Waveplate configuration in Fig.~\ref{fig:SM_fig_MP.jpg} for mutually unbiased bases $\{M_k\}_{k=0}^{d}$ in dimension $d=4$.}
\label{tab:waveplate-settings}
\resizebox{\textwidth}{!}{%
\begin{tabular}{|c|c|c|c|c|c|c|c|c|c|c|}
\hline
Waveplates
& \multicolumn{2}{c|}{$M_0$}
& \multicolumn{2}{c|}{$M_1$}
& \multicolumn{2}{c|}{$M_2$}
& \multicolumn{2}{c|}{$M_3$}
& \multicolumn{2}{c|}{$M_4$} \\
\hline
2-D Subspace & a & b
& a & b
& a & b
& a & b
& a & b \\
\hline 
Waveplate 1 & / & $\HWP(+\frac{\pi}{4})$ & $\HWP(-\frac{\pi}{8})$ & $\HWP(+\frac{\pi}{8})$ & $\HWP(-\frac{\pi}{8})$ & $\HWP(+\frac{\pi}{8})$ & $\QWP(-\frac{\pi}{4})$ & $\QWP(+\frac{\pi}{4})$ & $\QWP(-\frac{\pi}{4})$ & $\QWP(+\frac{\pi}{4})$ \\
\hline
Waveplate 2 & $\HWP(+\frac{\pi}{4})$ & / & $\HWP(+\frac{\pi}{8})$ & $\HWP(-\frac{\pi}{8})$ & $\HWP(-\frac{\pi}{8})$ & $\HWP(+\frac{\pi}{8})$ & $\QWP(+\frac{\pi}{4})$ & $\QWP(-\frac{\pi}{4})$ & $\QWP(-\frac{\pi}{4})$ & $\QWP(+\frac{\pi}{4})$ \\
\hline
Waveplate 3 & \multicolumn{2}{c|}{/} & \multicolumn{2}{c|}{$\HWP(+\frac{\pi}{8})$} & \multicolumn{2}{c|}{$\QWP(+\frac{\pi}{4})$}
& \multicolumn{2}{c|}{$\HWP(+\frac{\pi}{8})$}
& \multicolumn{2}{c|}{$\QWP(+\frac{\pi}{4})$} \\
\hline
\end{tabular}
}
\end{table}

\subsection{Maximum-likelihood estimation of the marginals}
\label{sec:mle_marginals}

In this section, we describe how the two-body marginals are reconstructed from the experimental QOT data. 
All density matrices shown in the main text are obtained using maximum-likelihood estimation (MLE) \cite{PhysRevA.64.052312}. 
Direct linear inversion of the count data may yield a nonphysical density matrix because of statistical fluctuations. 
MLE avoids this problem by searching over physical density matrices and selecting the one most consistent with the observed counts.

Let $X$ denote one of the six two-body subsystems,
\begin{equation}
    X\in\{AB,AC,AD,BC,BD,CD\}.
\end{equation}
The dimension of $X$ is
\begin{equation}
    D_X=d_{X_1}d_{X_2},
\end{equation}
where $(d_A,d_B,d_C,d_D)=(4,4,2,2)$. 
Thus $D_{AB}=16$, $D_{AC}=D_{AD}=D_{BC}=D_{BD}=8$, and $D_{CD}=4$.

\begin{figure}[h]
    \centering
    \includegraphics[width=1\linewidth]{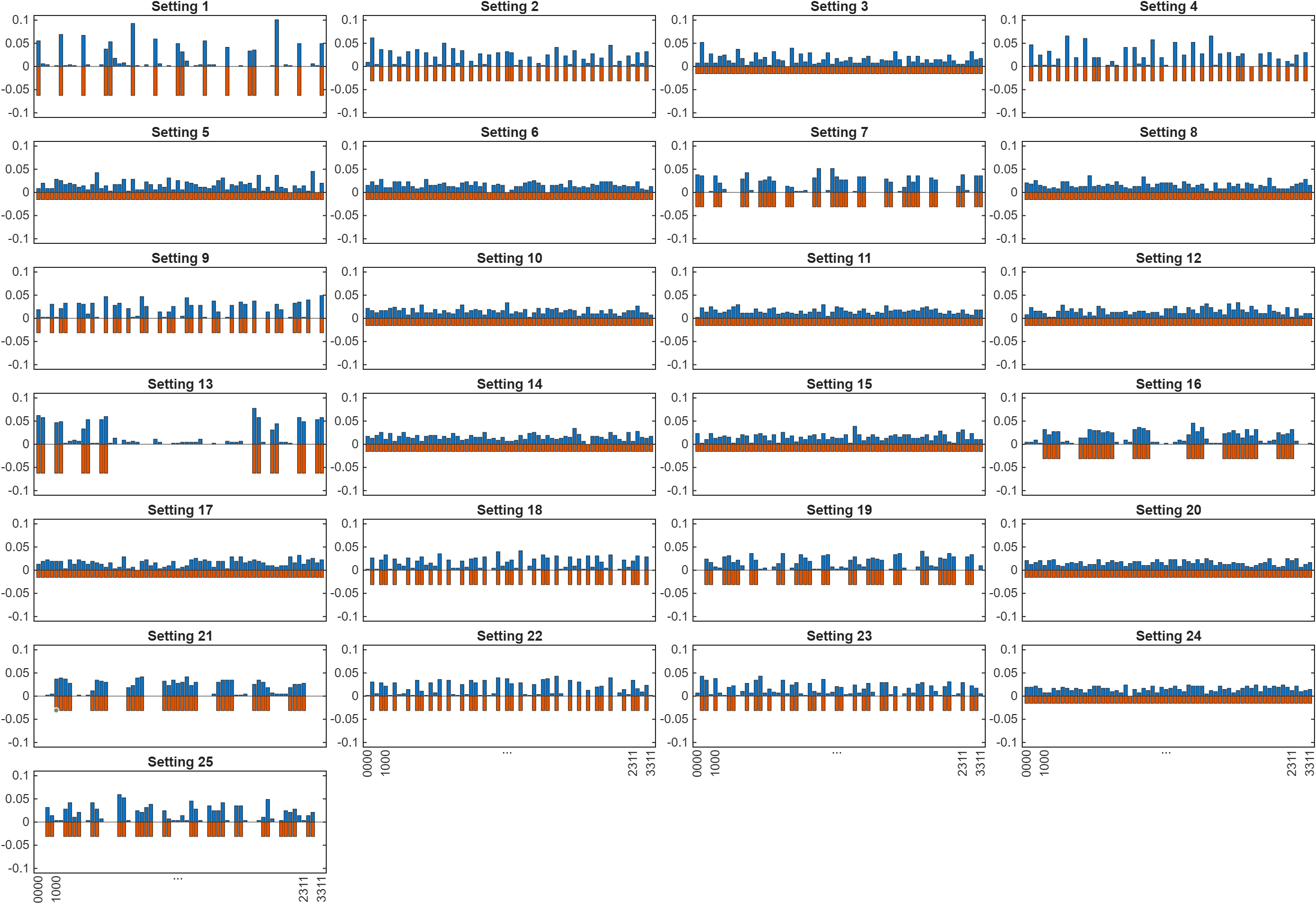}
    \caption{Normalized detected probabilities of the experimental state (blue bars) and expected probabilities for the target state (orange bars) for each measurement setting of high-dimensional QOT.}
    \label{fig:SM_fig_prob.jpg}
\end{figure}

For each global QOT setting $\alpha$, the experiment records fourfold coincidence counts for all $4\times 4\times 2 \times 2=64$ output combinations of the four parties. 
We denote these raw counts by
\begin{equation}
    N^{(o_A,o_B,o_C,o_D)}_{\alpha},
\end{equation}
where $o_i$ labels the local detector outcome of party $i$. The corresponding normalized probabilities are shown in Fig.~\ref{fig:SM_fig_prob.jpg}.

For a two-body subsystem $X=\{X_1,X_2\}$, the counts associated with local outcomes $(o_{X_1},o_{X_2})$ are obtained by summing over the outcomes of the complementary parties,
\begin{equation}
    N^{(o_{X_1},o_{X_2})}_{\alpha,X}
    =
    \sum_{\{o_j:j\notin X\}}
    N^{(o_A,o_B,o_C,o_D)}_{\alpha}.
\end{equation}
The total number of effective counts for subsystem $X$ under setting $\alpha$ is
\begin{equation}
    N_{\alpha,X}
    =
    \sum_{o_{X_1},o_{X_2}}
    N^{(o_{X_1},o_{X_2})}_{\alpha,X}.
\end{equation}

Let $\Pi^{(o_{X_1},o_{X_2})}_{\alpha,X}$ be the corresponding projective measurement operator on the two-body subsystem $X$,
\begin{equation}
    \Pi^{(o_{X_1},o_{X_2})}_{\alpha,X}
    =
    \ket{\psi^{(X_1)}_{\alpha,o_{X_1}}}
    \bra{\psi^{(X_1)}_{\alpha,o_{X_1}}}
    \otimes
    \ket{\psi^{(X_2)}_{\alpha,o_{X_2}}}
    \bra{\psi^{(X_2)}_{\alpha,o_{X_2}}}.
\end{equation}
For a trial density matrix $\rho_X$, the predicted probability for this outcome is
\begin{equation}
    p^{(o_{X_1},o_{X_2})}_{\alpha,X}(\rho_X)
    =
    \mathrm{Tr}
    \left[
        \rho_X
        \Pi^{(o_{X_1},o_{X_2})}_{\alpha,X}
    \right].
\end{equation}
The corresponding expected number of counts is
\begin{equation}
    \widetilde{N}^{(o_{X_1},o_{X_2})}_{\alpha,X}
    =
    N_{\alpha,X}
    \,
    p^{(o_{X_1},o_{X_2})}_{\alpha,X}(\rho_X).
\end{equation}

Following the standard MLE treatment of photonic quantum state tomography \cite{PhysRevA.64.052312}, we approximate the counting statistics by a Gaussian distribution in the large-count limit. 
The probability of observing the count
$N^{(o_{X_1},o_{X_2})}_{\alpha,X}$
given the trial state $\rho_X$ is therefore proportional to
\begin{equation}
    p\!\left(
        N^{(o_{X_1},o_{X_2})}_{\alpha,X};
        \rho_X
    \right)
    \propto
    \exp
    \left[
        -
        \frac{
        \left(
            N^{(o_{X_1},o_{X_2})}_{\alpha,X}
            -
            \widetilde{N}^{(o_{X_1},o_{X_2})}_{\alpha,X}
        \right)^2
        }{
        2\left(
            \sigma^{(o_{X_1},o_{X_2})}_{\alpha,X}
        \right)^2
        }
    \right],
\end{equation}
where the standard deviation is approximated by Poissonian counting noise,
\begin{equation}
    \sigma^{(o_{X_1},o_{X_2})}_{\alpha,X}
    \simeq
    \sqrt{
        \widetilde{N}^{(o_{X_1},o_{X_2})}_{\alpha,X}
    }.
\end{equation}

Assuming independent fluctuations for different outcomes and settings, the likelihood of obtaining the full set of counts for subsystem $X$ is
\begin{equation}
    p\!\left(
        \{N^{(o_{X_1},o_{X_2})}_{\alpha,X}\};
        \rho_X
    \right)
    =
    \frac{1}{\mathcal{N}}
    \prod_{\alpha,o_{X_1},o_{X_2}}
    \exp
    \left[
        -
        \frac{
        \left(
            N^{(o_{X_1},o_{X_2})}_{\alpha,X}
            -
            N_{\alpha,X}
            \mathrm{Tr}
            \left[
                \rho_X
                \Pi^{(o_{X_1},o_{X_2})}_{\alpha,X}
            \right]
        \right)^2
        }{
        2N_{\alpha,X}
        \mathrm{Tr}
        \left[
            \rho_X
            \Pi^{(o_{X_1},o_{X_2})}_{\alpha,X}
        \right]
        }
    \right],
    \label{eq:likelihood_product}
\end{equation}
where $\mathcal{N}$ is a normalization constant independent of $\rho_X$.

The MLE state is defined as the physical density matrix that maximizes this likelihood,
\begin{equation}
    \rho_X^{\mathrm{MLE}}
    =
    \arg\max_{\rho_X\in\mathcal{S}_X}
    p\!\left(
        \{N^{(o_{X_1},o_{X_2})}_{\alpha,X}\};
        \rho_X
    \right),
    \label{eq:MLE_argmax}
\end{equation}
where $\mathcal{S}_X$ denotes the set of positive semidefinite, unit-trace density matrices on subsystem $X$. 
Equivalently, maximizing Eq.~\eqref{eq:likelihood_product} is equivalent to minimizing the cost function
\begin{equation}
    \mathcal{L}_X(\rho_X)
    =
    \frac{1}{2}
    \sum_{\alpha,o_{X_1},o_{X_2}}
    \frac{
    \left(
        N^{(o_{X_1},o_{X_2})}_{\alpha,X}
        -
        N_{\alpha,X}
        \mathrm{Tr}
        \left[
            \rho_X
            \Pi^{(o_{X_1},o_{X_2})}_{\alpha,X}
        \right]
    \right)^2
    }{
    N_{\alpha,X}
    \mathrm{Tr}
    \left[
        \rho_X
        \Pi^{(o_{X_1},o_{X_2})}_{\alpha,X}
    \right]
    } .
\end{equation}

\begin{figure}[ht]
    \centering
    \includegraphics[width=1\linewidth]{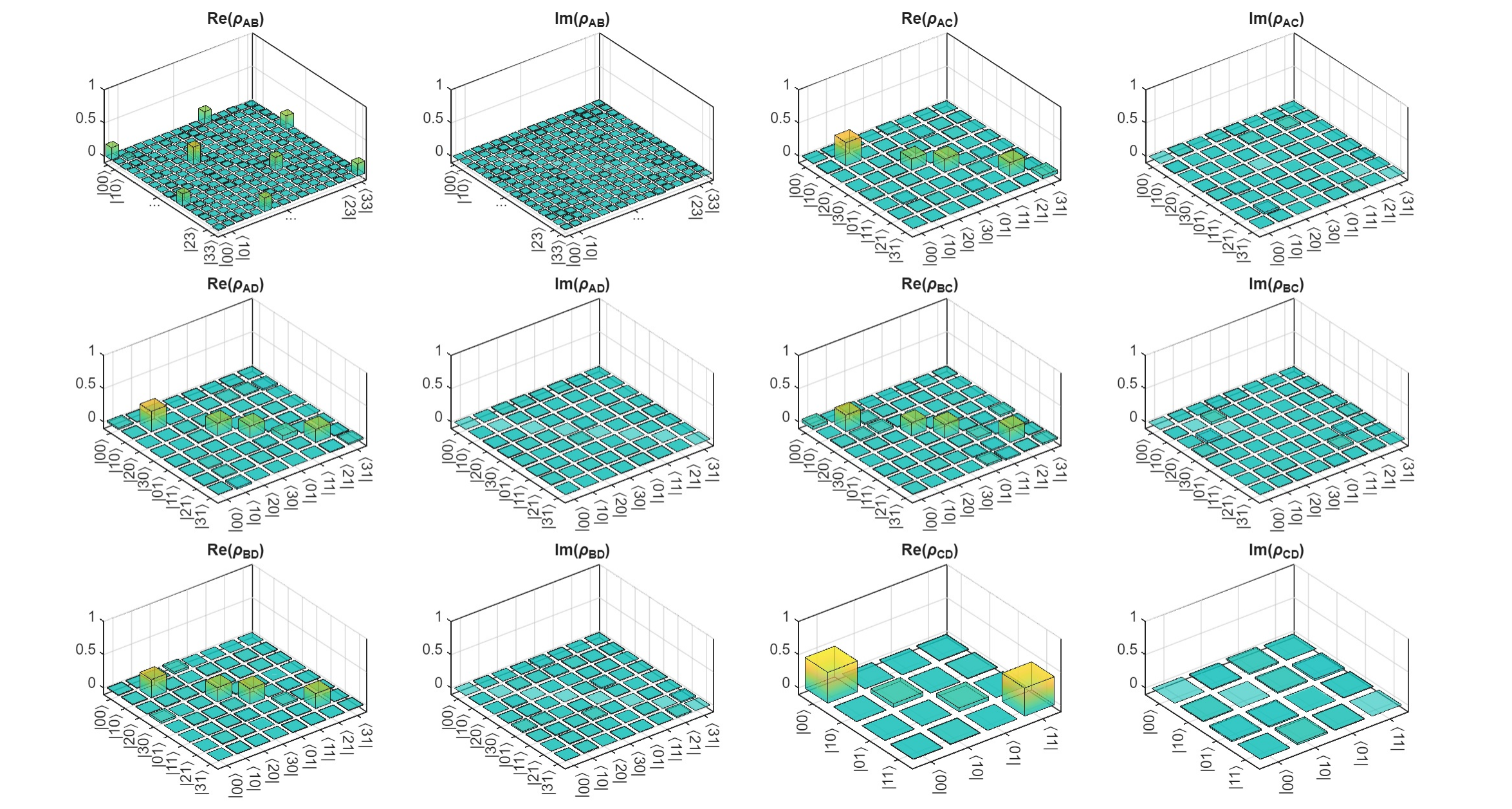}
    \caption{Real and imaginary parts of the experimental two-body marginals reconstructed by QOT.}
    \label{fig:SM_fig_marginals.jpg}
\end{figure}

Next, we parameterize the density matrix by a Cholesky decomposition,
\begin{equation}
    \rho_X(\mathbf{t})
    =
    \frac{
        T_X^\dagger(\mathbf{t})T_X(\mathbf{t})
    }{
        \mathrm{Tr}
        \left[
            T_X^\dagger(\mathbf{t})T_X(\mathbf{t})
        \right]
    },
\end{equation}
where $T_X(\mathbf{t})$ is a complex lower-triangular $D_X\times D_X$ matrix with real diagonal entries. 
This parametrization guarantees that $\rho_X(\mathbf{t})$ is Hermitian, positive semidefinite, and unit-trace for arbitrary real parameters $\mathbf{t}$, thereby reducing the constrained optimization problem in Eq.~\eqref{eq:MLE_argmax} to an unconstrained optimization problem:
\begin{equation}
    \mathbf{t}_{\mathrm{opt}}
    =
    \arg\min_{\mathbf{t}}
    \mathcal{L}_X\!\left[\rho_X(\mathbf{t})\right],
    \qquad
    \rho_X^{\mathrm{MLE}}
    =
    \rho_X(\mathbf{t}_{\mathrm{opt}}).
\end{equation}
This procedure is applied to all six two-body subsystems. 
The reconstructed physical density matrices are shown in Fig.~\ref{fig:SM_fig_marginals.jpg} and are used to compute the marginal fidelities, negativities, and global-to-local overlap ratios reported in the main text.

\subsection{Noise resilience of the inner-product criterion for entanglement dimensionality certification}

Here we compare the inner-product criterion (IPC) with the fidelity-based criterion (FBC) for certifying high-dimensional entanglement under added white noise. We use the Hilbert-Schmidt inner product notation $\langle A,B\rangle=\mathrm{Tr}(AB)$, and denote the experimental state by $\rho$ and the target state by $\sigma=\ket{\Lambda}\bra{\Lambda}$.
The global overlap is equal to the fidelity $\langle\rho,\sigma\rangle=\langle\Lambda|\rho|\Lambda\rangle$.

The FBC uses the fidelity witness for the Schmidt number.
For a pure target state with Schmidt decomposition
\begin{equation}
    \ket{\Psi}
    =
    \sum_{j}\sqrt{\lambda_j}\ket{e_j}_X\ket{f_j}_{\bar X},
    \qquad
    \lambda_1\geq \lambda_2\geq\cdots ,
\end{equation}
any state $\varrho$ with Schmidt number at most $r$ across $X|\bar X$ satisfies
\begin{equation}
    F
    =
    \bra{\Psi}\varrho\ket{\Psi}
    \leq
    \sum_{j=1}^{r}\lambda_j .
    \label{eq:SM_FBC}
\end{equation}
Thus, violating Eq.~\eqref{eq:SM_FBC} certifies a Schmidt number larger than $r$.
For the four bipartitions shown in Fig.~4(b) of the main text, namely
$A|BCD$, $B|ACD$, $AC|BD$, and $AD|BC$, the target state has four equal Schmidt coefficients $\lambda_j=1/4$.
Therefore, the $3$-FBC threshold is
\begin{equation}
    F>\frac{3}{4},
\end{equation}
which certifies Schmidt number larger than three.

IPC uses the global overlap together with the corresponding local overlaps.
For a bipartition $X|\bar X$, it defines the global-to-local overlap ratio:
\begin{equation}
    S_{X|\bar X}(\rho,\sigma)
    =
    \max
    \left\{
    \frac{\langle\rho,\sigma\rangle}
         {\langle\rho_X,\sigma_X\rangle},
    \frac{\langle\rho,\sigma\rangle}
         {\langle\rho_{\bar X},\sigma_{\bar X}\rangle}
    \right\},
    \label{eq:SM_IPC_ratio}
\end{equation}
where $\rho_X=\mathrm{Tr}_{\bar X}(\rho)$, $\sigma_X=\mathrm{Tr}_{\bar X}(\sigma)$, and similarly for $\bar X$.
The IPC condition $S_{X|\bar X}(\rho,\sigma)>r$ certifies that the Schmidt number across $X|\bar X$ is larger than $r$ \cite{li2026simultaneous, PRXQuantum.2.040357}.
For the four bipartitions with target Schmidt rank four, this gives the $3$-IPC condition
\begin{equation}
    S_{X|\bar X}>3.
\end{equation}

\begin{figure}[ht]
    \centering
    \includegraphics[width=.5\linewidth]{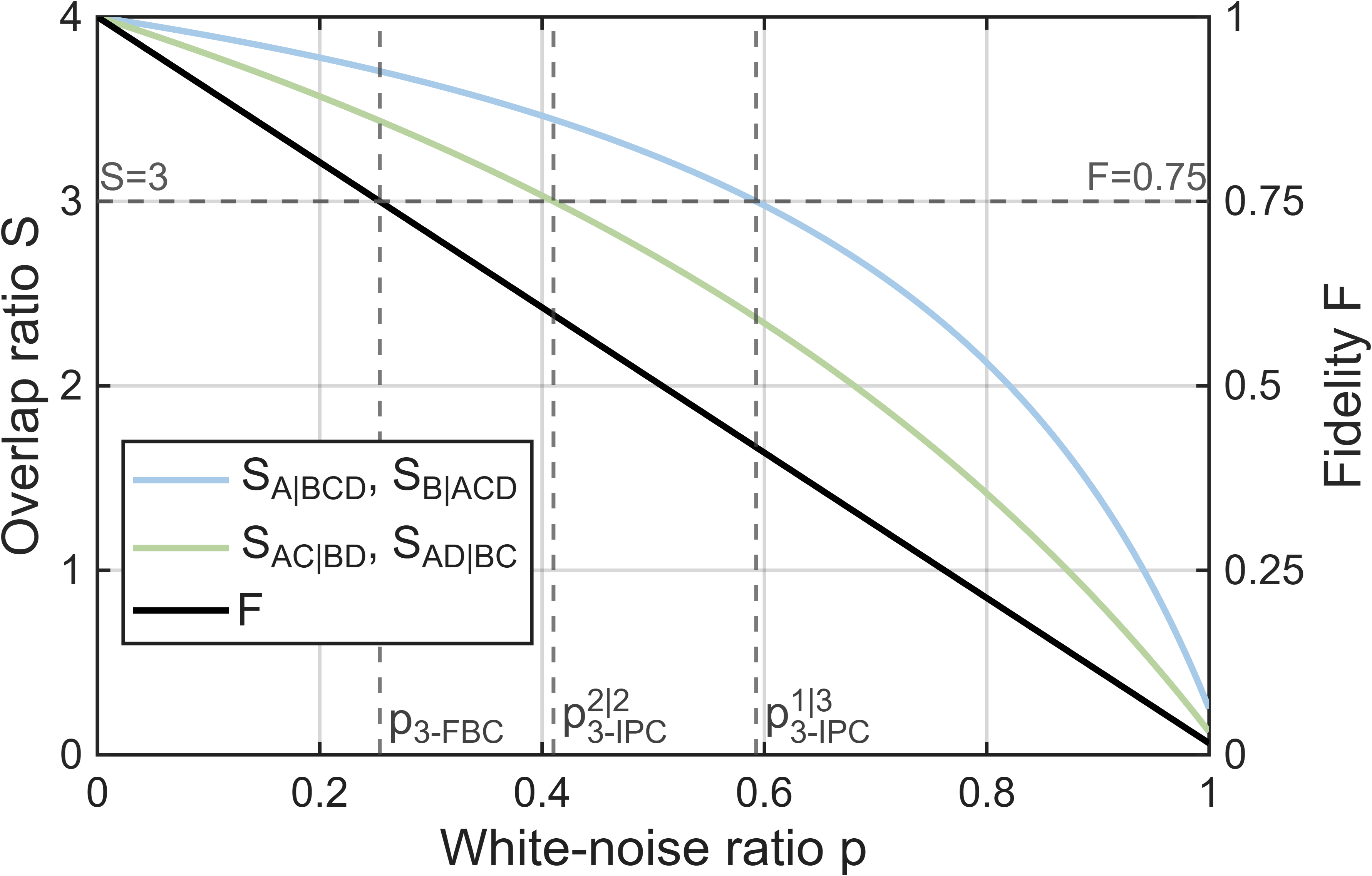}
    \caption{Comparison of the Inner-Product Criterion (IPC) and the Fidelity-Based Criterion (FBC) for the depolarized target state in Eq.~\eqref{eq:depol}. The 3-IPC shows a clear advantage over the 3-FBC as $S_{X|\bar{X}}$ can remain above threshold when $F$ falls below threshold.
    }
    \label{fig:SM_fig_IPC.png}
\end{figure}

To illustrate the stronger detection power of IPC than that of FBC in our case, we consider the depolarized target state
\begin{equation}
    \rho(p) = (1-p)|\Lambda\rangle\langle\Lambda| + p \frac{\mathbb{I}_D}{D},
    \label{eq:depol}
\end{equation}
where $p$ is the ratio of the white noise, $D=d_Ad_Bd_Cd_D=4\times4\times2\times2=64$ is the global dimension and $\mathbb{I}_D$ is the identity operator in the $D$-dimensional Hilbert space. As shown in Fig.~\ref{fig:SM_fig_IPC.png}, the 3-IPC has a clear advantage over the 3-FBC. The fidelity falls below threshold at white noise $p_{3-\mathrm{FBC}}=0.254$, whereas the overlap ratios for the $2|2$ cuts and $1|3$ cuts fall below threshold at larger white noise $p_{3-\mathrm{IPC}}^{2|2}=0.410$ and $p_{3-\mathrm{IPC}}^{1|3}$${}=0.593$, respectively. Between $p_{3-\mathrm{FBC}}$ and $p_{3-\mathrm{IPC}}$ are the $(3+1)$-unfaithful states \cite{li2026simultaneous} which are detectable by 3-IPC but not by 3-FBC.

\subsection{Data postprocessing for the experimental noisy states}
To investigate noise resilience, we recorded coincidence counts generated by independent white-noise sources coupled into the detectors. Several noise levels were obtained by varying the optical power of these sources, using the same measurement settings and integration times as for the original experiment. 
The measured noise counts were then superposed with the original coincidence counts, producing noisy experimental datasets with different effective noise parameters $p$. The fidelities, overlap ratios, and their uncertainties were evaluated from the combined counts using the same analysis pipeline as for the original data, producing the data points in Fig.~4(b) in the main text.

The curves in Fig.~4(b) in the main text are extracted from the depolarizing channel acting on the experimental state $\rho_{\exp}$:
\begin{equation}
    \rho_{\exp}(p)=(1-p)\rho_{\exp}+p\frac{\mathbb{I}_D}{D},
\end{equation}
then the global fidelity and the local overlaps transform as
\begin{equation}
    F(p)=(1-p)F_{\exp}+\frac{p}{D},
    \qquad
    O_X(p)=(1-p)O_X^{\exp}+\frac{p}{D_X},
    \label{eq:SM_depolarizing_overlaps}
\end{equation}
where $O_X=\langle \rho_X,\sigma_X\rangle$ for subsystem $X$ and $D_X$ is the dimension of the subsystem $X$. $F_{\exp}$ and $O_X^{\exp}$ are the experimental values without the additional white noise.
The overlap ratio curves are therefore induced as
\begin{equation}
    S_{X|\bar X}(p)
    =
    \max
    \left\{
    \frac{F(p)}{O_X(p)},
    \frac{F(p)}{O_{\bar X}(p)}
    \right\}.
    \label{eq:SM_depolarizing_IPC}
\end{equation}

\subsection{Origin of the fidelity deviations}
\label{sec:fidelity_deviations}

The marginal fidelities in Fig.~3(g) of the main text characterize the deviations of the prepared marginals from their ideal targets. Measurement errors are negligible because all waveplate angles were calibrated. The statistical uncertainties of 0.012--0.017 are much smaller than the deviations of 0.06--0.19. The deviations therefore originate mainly from state preparation. The global data of Tab.~\ref{tab:rho-real} show two imperfections. First, 10.9\% of the population lies outside the four target components. This population lowers the global fidelity and, to the extent that it falls outside the supports of the target marginals, the marginal fidelities. Second, the six measured coherences together reach 88\% of their upper bounds $\sqrt{\rho_{\mu\mu}\rho_{\nu\nu}}$ set by the measured populations. This coherence deficit lowers the global fidelity from 0.890 to 0.809.

Among the marginals, the coherence deficit affects only $\rho_{AB}$. This follows from the generation of the target state described in Sec.~S2\,A and in the main text. The first PBS fuses one photon from each of the two photon-pair sources and produces the two components $|HHHH\rangle$ and $|VVVV\rangle$ of the four-photon GHZ state. Imperfect interference at this PBS reduces the coherence between these two components. Since the two components differ in the state of every party, this coherence vanishes in all two-body marginals. Its loss lowers the global fidelity but none of the marginal fidelities. The second PBS, in the dimension-increasing device, creates the polarization coherences between $|HH\rangle$ and $|VV\rangle$ of the two ququart photons within each component, that is, $\langle 00|\rho_{AB}|22\rangle$ and $\langle 11|\rho_{AB}|33\rangle$ of the ideal $AB$ marginal. The other five ideal marginals are diagonal. Imperfect interference at the second PBS therefore lowers $F_{AB}$ but leaves the five diagonal marginals unaffected.

\bibliography{bib-SM}